\documentclass[12pt]{article}

\usepackage{newtxtext,newtxmath}

\usepackage{graphicx}

\usepackage[letterpaper,margin=1in]{geometry}

\renewenvironment{abstract}
	{\quotation}
	{\endquotation}

\date{}

\makeatletter
\renewcommand{\fnum@figure}{\textbf{Figure \thefigure}}
\renewcommand{\fnum@table}{\textbf{Table \thetable}}
\makeatother

\usepackage{scicite}

\usepackage{url}

\def\scititle{High-fidelity spatial information transfer through dynamic scattering media by an epsilon-near-zero time-gate}
\title{\bfseries \boldmath \scititle}

\author{
	Yang~Xu$^{1\ast\dagger}$,
	Saumya~Choudhary$^{2\dagger}$,
        Long~D.~Nguyen$^{1}$,
        Matthew~Klein$^{3,4}$,\and
        Shivashankar Vangala$^{4}$, J. Keith Miller$^{5}$,
        Eric G. Johnson$^{6}$, \and
        Joshua R. Hendrickson$^{4}$, M. Zahirul Alam$^{7}$,
	Robert~W.~Boyd$^{1,2,7}$\and
	\small$^{1}$Department of Physics and Astronomy, University of Rochester, Rochester, New York 14627, USA.\and
	\small$^{2}$The Institute of Optics, University of Rochester, Rochester,
New York 14627, USA.\and
        \small$^{3}$KBR, Beavercreek, Ohio 45431, USA.\and
        \small$^{4}$Sensors Directorate, Air Force Research Laboratory, Wright-Patterson AFB, Ohio 45433, USA.\and
        \small$^{5}$The Holcombe Department of Electrical and Computer Engineering, Clemson Center for Optical Materials Science \and \small and Engineering Technologies, Clemson, South Carolina 29634, USA.\and
        \small$^{6}$CREOL, The College of Optics and Photonics at the University of Central Florida, Orlando, Florida 32816, USA.\and
        \small$^{7}$Department of Physics, University of Ottawa, Ottawa, Ontario K1N 6N5, Canada.\and
	\small$^\ast$Corresponding author. Email: yxu100@ur.rochester.edu\and
	\small$^\dagger$These authors contributed equally to this work.
}

\begin{document} 

\maketitle

\begin{abstract} \bfseries \boldmath
Transparent conducting oxides (TCO) such as indium-tin-oxide (ITO) exhibit strong optical nonlinearity in the frequency range where their permittivities are near zero. We leverage this nonlinear optical response to realize a sub-picosecond time-gate based on upconversion (or sum-) four-wave mixing (FWM) between two ultrashort pulses centered at the epsilon-near-zero (ENZ) wavelength in a sub-micron-thick ITO film. The time-gate removes the effect of both static and dynamic scattering on the signal pulse by retaining only the ballistic photons of the pulse, that is, the photons that are not scattered. Thus, the spatial information encoded in either the intensity or the phase of the signal pulse can be preserved and transmitted with high fidelity through scattering media. Furthermore, in the presence of time-varying scattering, our time-gate can reduce the resulting scintillation by two orders of magnitude. In contrast to traditional bulk nonlinear materials, time gating by sum-FWM in a sub-wavelength-thick ENZ film can produce a scattering-free upconverted signal at a visible wavelength without sacrificing spatial resolution, which is usually limited by the phase-matching condition. Our proof-of-principle experiment can have implications for potential applications such as \textit{in vivo} diagnostic imaging and free-space optical communication.
\end{abstract}

\noindent
Unwanted scattering of light during propagation can severely degrade the quality of the received optical signal, thereby adversely affecting applications such as imaging, sensing, and communication. An ultrafast time-gate can undo the effect of scattering and recover the original unscrambled signal by producing a response only from the unaffected (or the ``ballistic") part of the signal. The photons comprising this ballistic signal exit the scattering medium the earliest as they experience no scattering events. The ability to select (or gate) only the ballistic photons is of particular importance to many noninvasive biomedical imaging techniques that are primarily limited by light scattering \cite{wang1991ballistic, linne2006ballistic, paciaroni2004single, ipus2019scope}. For instance, in transillumination imaging for breast cancer diagnostics, the light scattered by the breast tissue can severely degrade the quality of the shadow image of tumors \cite{bartrum1984transillumination, sickles1984breast, geslien1985transillumination, cutler1931transillumination, merritt1984real}. Previous demonstrations of optical time-gates to reject ballistic photons have mostly employed the optical Kerr effect \cite{duguay1969ultrafast, wang1991ballistic, wang1993time, das1993ultrafast, alfano1994time, d2012experimental}. In an optical Kerr gate, the nonlinear optical medium is placed between two crossed polarizers. As such, the ballistic part of the signal pulse carrying the image is transmitted only when it spatio-temporally overlaps with an intense linearly polarized gate pulse that rotates the polarization of the signal beam. However, even though the nonresonant electronic Kerr response is ultrafast, it is usually weak in most bulk nonlinear optical materials and requires long interaction lengths to build up a transient birefringence for a sufficient signal-to-noise ratio in the transmitted signal. This long interaction length limits the acceptance angle of the Kerr medium, which constrains the maximum achievable imaging resolution \cite{huang2022high}. Hence, despite several experimental attempts to improve the Kerr gating response \cite{yoo1991imaging, sedarsky2011quantitative, zheng2017ballistic, mathieu2014time, zhan2014optical, tan2013high, tan2014sharpness, huang2022high}, achieving good image contrast and robustness against strong scattering simultaneously remains a challenge for traditional Kerr media. 

Epsilon-near-zero (or ENZ) materials demonstrate exotic linear and nonlinear optical wave phenomena that manifest as a consequence of their vanishing permittivity \cite{liberal2017near, reshef2019nonlinear}. In particular, materials have been shown to exhibit a unity-order change in the nonlinear refractive index \cite{alam2016large, caspani2016enhanced, alam2018large}, enhanced harmonic generation \cite{wang2000second, capretti2015enhanced, capretti2015comparative, yang2019high}, frequency conversion \cite{zhou2020broadband, pang2021adiabatic, liu2021photon} and wave-mixing \cite{carnemolla2021visible}. These enhanced nonlinear optical processes have been typically demonstrated on platforms of transparent conducting oxides (or TCOs) such as indium-tin-oxide (ITO) or aluminum-doped zinc oxide (AZO). TCOs are degenerately-doped semiconductors with a large concentration of free electrons leading to permittivities that can be described by the Drude-Lorentz model \cite{khurgin2021fast}. Crucially, the nonlinear optical response in these materials has a sub-picosecond time scale due to the electron dynamics in a non-parabolic band structure \cite{khurgin2021fast}. This giant ultrafast nonlinearity, together with the tunability of their ENZ regime through doping, makes TCOs ideal platforms for realizing efficient optical switches, as well as ultrafast coherent time-gates to select a part of a signal within a very small (sub-ps) time window \cite{jaffray2022near}.

Here, we demonstrate, to our knowledge, the first ultrafast time-gate based on the upconversion (or sum-) four-wave mixing (FWM) process in a sub-wavelength-thick film of ITO \cite{naik2011oxides, alam2016large} for a scattering-free transmission of optical fields through static and dynamic scattering media. We show the preservation of the intensity and the phase distributions in the ballistic signal by recovering the spatial structure of an amplitude object and a beam carrying orbital angular momentum (OAM). The enhanced nonlinearity of ITO at ENZ due to the high field enhancement within the film \cite{reshef2017beyond, campione2013electric, caspani2016enhanced, kaipurath2016optically, lee2018strong, vassant2012berreman} and the relaxation of the phase-matching condition due to its sub-wavelength thickness yield a large sum-FWM signal. Thus, our ENZ time-gate offers a solution to the long-standing trade-off between the strength of the ballistic signal and the spatial resolution. Furthermore, the sum-FWM process, wherein two photons from the gate beam and one photon from the signal beam, all at the near-infrared (NIR) ENZ frequency, upconvert into a signal photon at a visible frequency, enables efficient signal collection with common silicon-based detectors.

\subsection*{The ultrafast FWM time-gate in ITO}
\begin{figure}
  \centering
  \includegraphics[width=\textwidth]{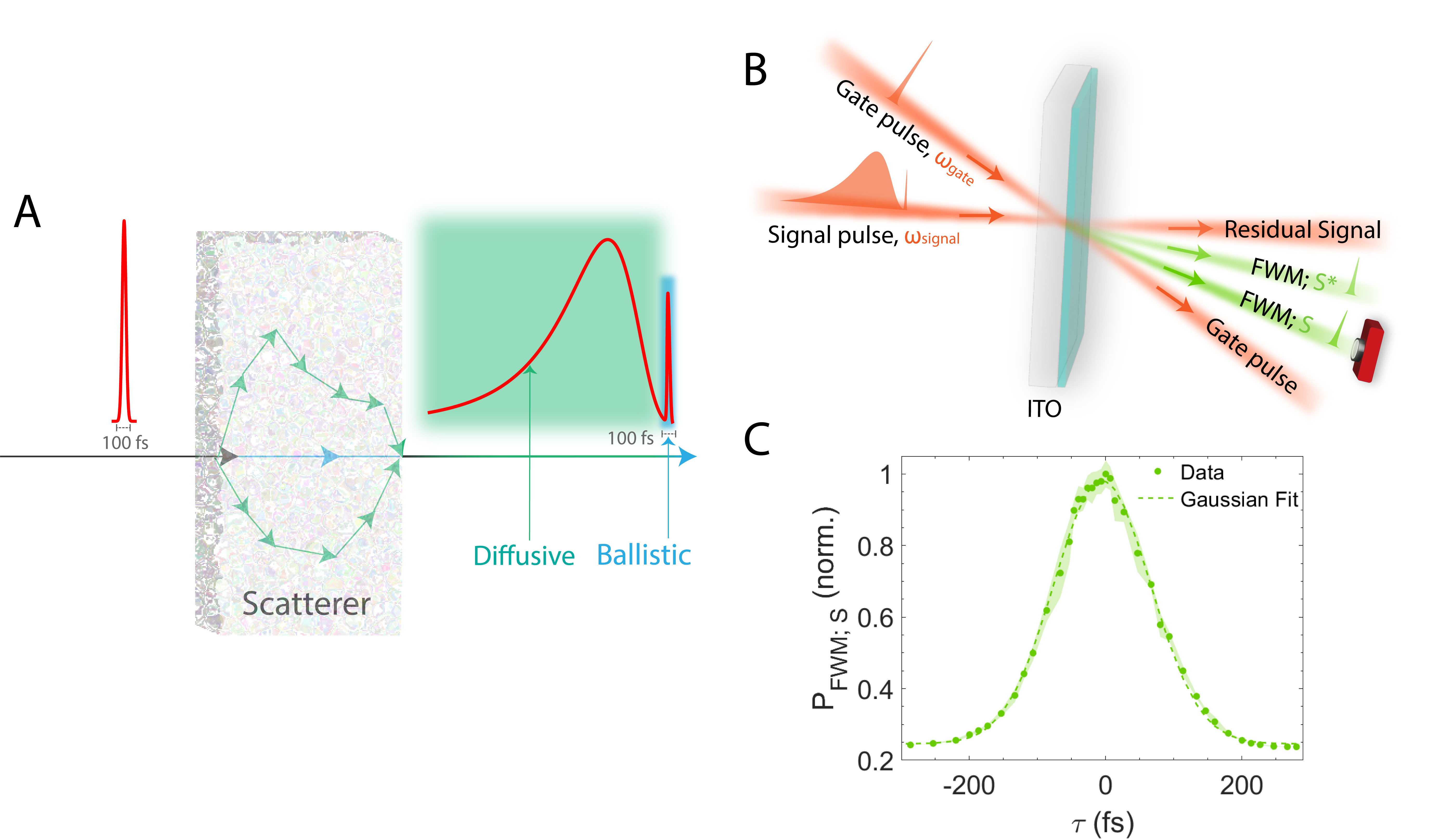}
\caption{\textbf{The selection of ballisic photons with a FWM time-gate.} \textbf{(A)} Diagram illustrating the ballistic and diffusive parts of the scattered pulse. A 100-fs-wide incident pulse experiences multiple scattering inside a medium. The ballistic portion of the pulse retains the shape of the incident pulse, while the diffusive photons constitute the elongated pulse that follows the ballistic peak. \textbf{(B)} The configuration of gate pulse and the scattered signal pulse for realizing the FWM gate. The sum-FWM (FWM; S) (drawn in green) is generated at the third-harmonic or $3\omega$ of the signal frequency $\omega$. \textbf{(C)} The normalized power $P$ of the sum-FWM signal at a wavelength of 517 nm as a function of the time delay $\tau$ between the gate and the signal pulses. The FWM response is essentially the autocorrelation between two 100-fs-wide Gaussian pulses \cite{jaffray2022near}, which demonstrates the fast nonlinear response associated with FWM in ITO.}

\label{fig:ito_gating}
\end{figure}

As shown in the diagram in Fig.~\ref{fig:ito_gating}A, when a light pulse propagates through a random scattering medium, the photons in the pulse could undergo none or several scattering events before they exit the medium \cite{jonsson2020multi, frantz2022multi}. The spatial and temporal distribution of the field as well as the spatial coherence of the ballistic photons, which experience no scattering events and exit the medium at the earliest, remain unchanged. On the contrary, the diffusive photons, experiencing one or more scattering events, spend a longer time inside the scattering medium, which degrades their spatial coherence \cite{pierrat2005spatial, alfano1994time}. As a consequence of light scattering, the temporal profile of the pulse on exiting the medium, as shown in Fig.~\ref{fig:ito_gating}A, consists of a small ballistic peak that is an attenuated replica of the incident pulse, followed by a large and broad diffusive component (see supplementary text for more details). Hence, selecting only this ballistic peak through a time-gate should enable the recovery of the original spatial and temporal information carried by the incident signal, except in situations where the scattering is strong enough to deplete the ballistic photons. 

We employ the sum-FWM process in a non-collinear geometry between the ballistic peak of an information-carrying signal pulse and an intense gate pulse. Figure~\ref{fig:ito_gating}B shows an illustration of the concept, and Fig.~\ref{fig:ito_gating}C shows the efficiency of the sum-FWM process with temporal overlap between the signal and gate pulses. The spatial separation of the sum-FWM pulse with the scattered signal pulse at the detector guarantees a complete rejection of the scattering background. Both the gate pulse and the unscattered signal pulse are 100-fs-long and centered at a wavelength of 1550 nm, which is close to the ENZ wavelength of the ITO film at 1515 nm (see Fig. \ref{fig:ito_epsilon} for the permittivity data of the ITO used). The wavevectors of the FWM responses obtained from momentum conservation are given by $\Vec{\boldsymbol{k}}_{\rm FWM; \, S} = 2\Vec{\boldsymbol{k}}_{\rm gate} + \Vec{\boldsymbol{k}}_{\rm signal}$ for the sum-FWM signal at the frequency $\omega_{\rm FWM; \, S} = 2\omega_{\rm gate} + \omega_{\rm signal}$ (see Fig. \ref{fig:fwm_ito}A in the supplementary materials for the energy level diagrams of all possible FWM processes that occur in ITO). Thus, the sum-FWM signal is centered at a wavelength of 517 nm (or the frequency-triple of the gate and the signal pulse) and is emitted at a smaller transverse angle relative to the incident pulses. This smaller emission angle of the sum-FWM pulse also reduces its spatial walk-off with the gate and signal pulses, thereby also reducing any potential resolution loss in imaging. In addition, we note that the nonlinear polarization for the sum-FWM process is $P_{\rm NL, FWM; S} = 3\epsilon_0 \chi^{(3)} \boldsymbol{E}^2_{\rm gate} \boldsymbol{E}_{\rm signal}$. Thus $P_{\rm NL, FWM; S}$ is directly proportional to the signal field ($\boldsymbol{E}_{\rm signal}$) itself, which further justifies our use of the sum-FWM response to recover the spatial information carried by the signal field.

\subsection*{Upconversion imaging through scattering media}

\begin{figure}
  \centering
  \includegraphics[width=\textwidth]{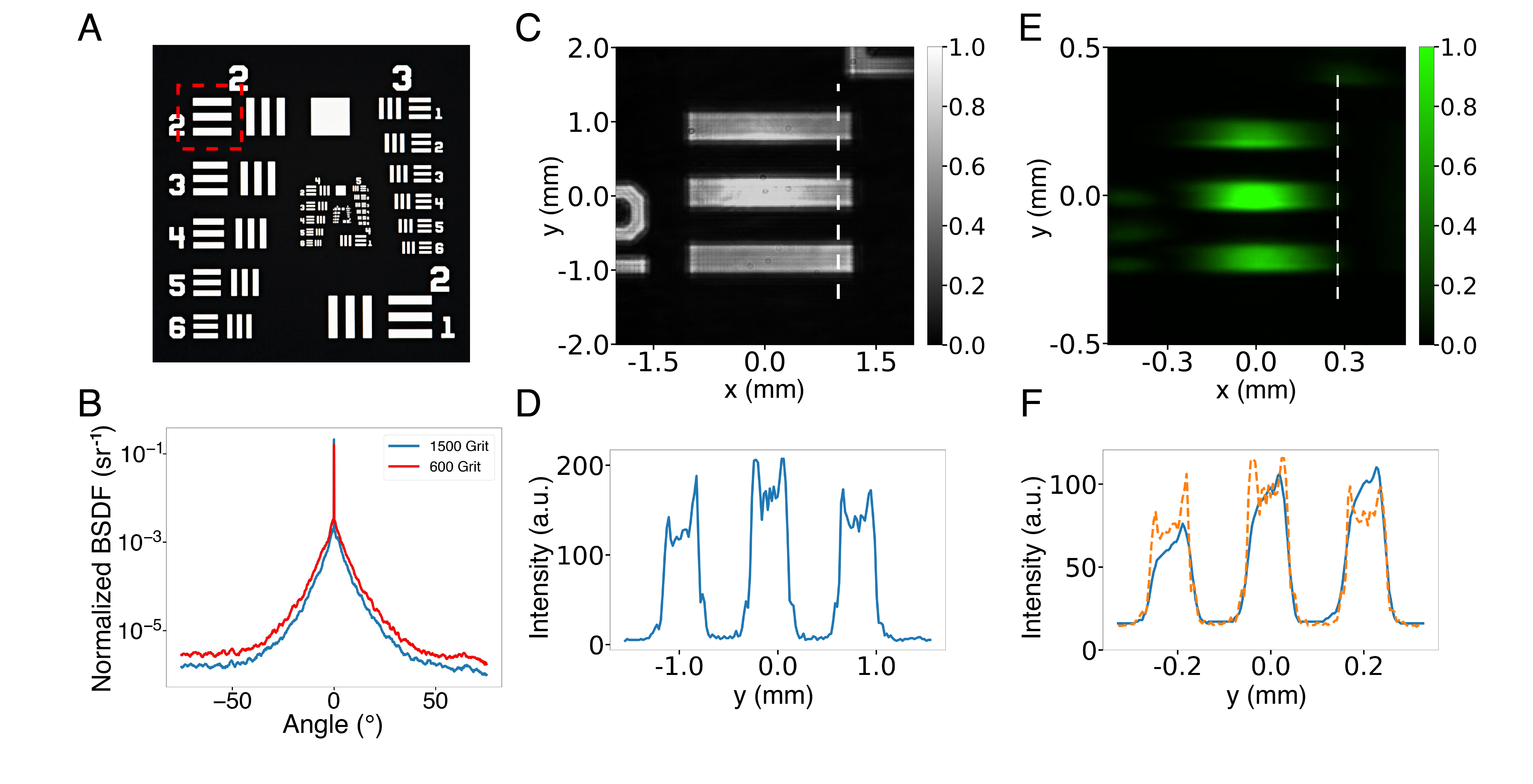}
\caption{\textbf{Characteristics of the scattering-free imaging system and the ground glass diffusers.} \textbf{(A)} A negative USAF 1951 resolution chart with the object boxed by the red dashed line. \textbf{(B)} The bidirectional scattering distribution function (BSDF) of the 1500-grit (blue) and 600-grit (red) ground glass optical diffusers. \textbf{(C)} The image of the target \textbf{(D)} and its line cut along the white dashed line in the signal field without any scattering medium in place. The corresponding \textbf{(E)} image and its \textbf{(F)} line cut profile of the sum-FWM field. The line cut of the undistorted signal image (orange dashed line) is rescaled to the same size as the FWM signal for comparing imaging quality. The green colormap has been used in the sum-FWM images for clarity.}
\label{fig:info_images}
\end{figure}

We first demonstrate the use of our time-gate for the upconversion imaging of an amplitude-only object through strong static scattering media. The test object is a set of three equally spaced horizontal bars on a negative USAF 1951 resolution target (highlighted in red in Fig.~\ref{fig:info_images}A). The width of each bar, as well as the spacing between two adjacent bars is 200 $\mu \text{m}$. We position the scatterer directly after the object in the path of the signal pulse. Figures~\ref{fig:info_images}C and \ref{fig:info_images}E show the amplitude images of the unscattered signal field and the sum-FWM field, respectively, along with their corresponding line cuts in the adjacent panels below. We note that the sum-FWM image is nearly $4\times$ smaller than the direct image of the object in the signal field for two reasons: (1) the degenerate sum-FWM is proportional to the cube of the fundamental field, which yields a focal spot size that is $3\times$ smaller than the fundamental field, and (2) the collimating lens in the path of the signal (sum-FWM) pulse has a focal length of 50 (40) mm, which further adds a factor of $1.25\times$ in the difference in magnification. We also note that despite the $3\times$ smaller focal spot of the sum-FWM field, there is no overall difference in imaging resolution between the direct image of the scattered signal and the sum-FWM image as the wavelength for sum-FWM is also reduced by a factor of 3.

\begin{figure}
  \centering
  \includegraphics[width=0.7\textwidth]{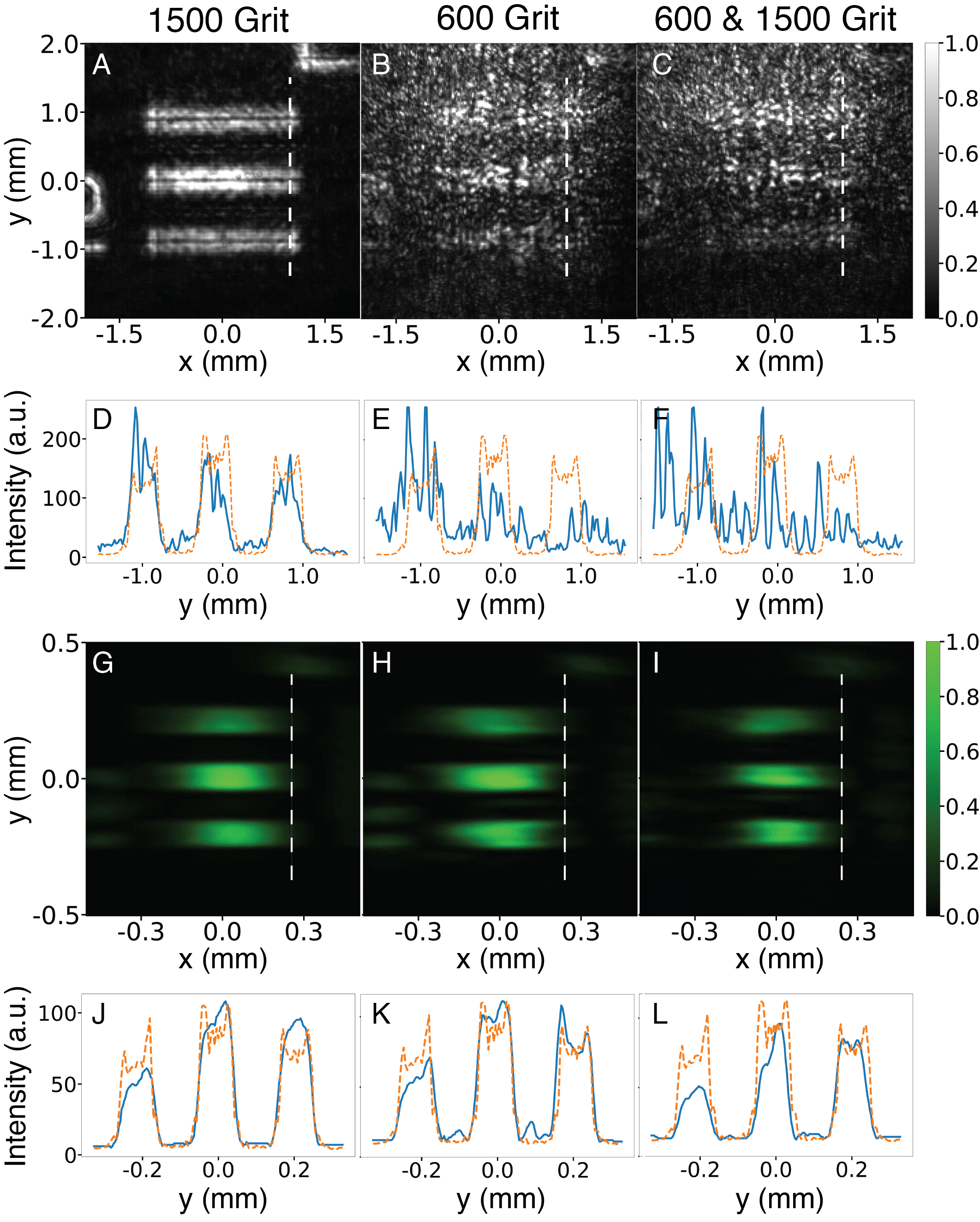}
\caption{\textbf{Scattering-free upconversion imaging through ground glass diffusers using sum-FWM in an ITO film. } Images of the signal field with diffusers of grit numbers of \textbf{(A)} 1500, \textbf{(B)} 600, and \textbf{(C)} 1500 and 600 combined. The panels \textbf{(D)-(F)} show the corresponding line cuts (solid, blue) along the white dashed line. The line cut of the undistorted signal image from Fig. \ref{fig:info_images}F (dashed, orange) is overlaid for comparison. The panels \textbf{(G)-(I)} show the images of the sum-FWM field for the signal images in panels \textbf{(A)-(C)} respectively, and the corresponding line cuts \textbf{(J)-(L)} along the dashed white line. The line cut of the unscattered signal image rescaled to the CMOS sensor (dashed, orange) is overlaid for comparison. Both the image resolution and the signal-to-noise contrast are well preserved in the sum-FWM output. }
\label{fig:usaf_images}
\end{figure}

We first use a set of two 2-mm-thick ground glass optical diffusers with grit numbers of 1500 and 600 as the static scatterers. From the bidirectional scattering distribution function (BSDF) shown in Fig. \ref{fig:info_images}B that characterizes the scattering strength of the diffusers, we note that the outgoing light scattered by the 600-grit ground glass optical diffuser has a wider angular distribution, thereby indicating a larger scattering strength. Figures~\ref{fig:usaf_images}A-C show the direct images of the signal field, and Figs.~\ref{fig:usaf_images}G-I the corresponding images of the sum-FWM field with the 1500-grit diffuser, the 600-grit diffuser, and the 1500- and 600-grit diffusers combined, respectively. Figures \ref{fig:usaf_images}D-F and \ref{fig:usaf_images}J-L, show the associated line cuts along the dashed white line in the direct images (Figs. \ref{fig:usaf_images}A-C) and the sum-FWM images (Figs. \ref{fig:usaf_images}G-I). We varied the power of the signal pulse for the different diffusers to overcome the corresponding power extinction and to maintain similar exposure times on the CMOS camera used to capture the sum-FWM images. Increasing the scattering strength of the diffuser leads to a progressive degradation of the image quality in the scattered signal to the extent that the object becomes indiscernible when the two diffusers are combined. In contrast, the structure of the object stays well preserved in the sum-FWM image for all three scattering strengths, and the generated sum-FWM field is bright enough for a commercially available CMOS camera (Thorlabs Zelux 1.6 MP monochrome CMOS camera) to detect. The contrast in quality between the scattered signal and the sum-FWM images is more evident in the respective line cuts, wherein the features of the object are completely submerged by the scattering noise in panels B and C, whereas a marginal difference between panels G, H and I is present for the sum-FWM image. 

We quantify the image quality for the various scatterers considered using the peak signal-to-noise ratio (PSNR) \cite{korhonen2012peak} and the Pearson correlation coefficient (PCC) \cite{cohen2009pearson}. PSNR is a common metric to quantify the strength of the obscuring noise that affects the image quality \cite{yuanji2003image, poobathy2014edge}. PCC gives a linear correlation between the pixel array of the reference image and the pixel array of the examined image and is often used to describe the similarity of the patterns in two images (see Materials and Methods). Table \ref{tab:psnr_pcc_images} shows the PSNR and PCC metrics of the direct images of the scattered signal and sum-FWM images for the various diffusers considered. We use the signal image (Fig. \ref{fig:info_images}C) and sum-FWM image (Fig. \ref{fig:info_images}E) taken without any scatterer as the reference images to exclude the effects of aberrations and the intensity profile of the illuminating signal beam. We note from Table~\ref{tab:psnr_pcc_images} that the PSNR drops by 6.47 dB, and the PCC is reduced by a factor of 4 for the signal image from the lowest to the highest scattering strength, whereas both metrics remain mostly constant for the sum-FWM image. We note that the slightly higher PSNR of the sum-FWM image for the 600-grit diffuser could be attributed to a relatively higher signal power, which was set by a visual inspection of the sum-FWM image contrast, leading to a slightly brighter sum-FWM image than for the other two diffusers. These metrics align with the result in Fig. \ref{fig:usaf_images}C where the object is completely obscured in the signal image for the highest scattering strength but remains intact in the sum-FWM image in Fig. \ref{fig:usaf_images}I.

\subsection*{Transmission of phase structures through
scattering media}
The recovery of the underlying signal by time gating the ballistic photons applies not only to its amplitude but also to its phase. Hence, we show that the phase distribution of a spatially structured beam, specifically a beam carrying an orbital angular momentum (OAM) \cite{yao2011orbital, oam_entangle}, is also well preserved in the gated sum-FWM field even in the presence of relatively strong scattering. In recent years, the OAM of light has become a promising candidate for free-space quantum communication \cite{willner2021orbital} because of its high information capacity \cite{pang2018demonstration} and its compatibility with other degrees of freedom, such as polarization \cite{nemirovsky2024increasing}. However, turbulence in free-space links inevitably leads to strong intermodal crosstalk that severely distorts the received signal \cite{zhou2021high, hamadou2013orbital}. Therefore, the transfer of OAM through turbulent media with high fidelity is a crucial step in the development of robust OAM-based free-space communication protocols \cite{li2018atmospheric, li2019mitigation}. 

In this demonstration (see Materials and Methods), we swap the imaging system in the path of the signal pulse with a spiral phase plate that introduces an OAM of topological charge $l = 2$ in the signal beam. The generated sum-FWM field is given by

\begin{align}
    E_{\rm FWM; \, S}(\mathbf{r}, 2\omega_{\text{gate}} + \omega_{\text{signal}}) \propto \left[E_{\rm gate}(\mathbf{r}, \omega_{\text{gate}})\right]^2 E_l(\mathbf{r}, \omega_{\text{signal}}),
\end{align}
where $E_{\rm gate}(\mathbf{r}, \omega_{\text{gate}})$ is approximately Gaussian. $E_l(\mathbf{r}, \omega_{\text{signal}})$ is the signal field after the spiral phase plate and carries a topological charge $l$ \cite{kotlyar2005generation}, so the sum-FWM field $E_{\rm FWM; S}(\mathbf{r}, 2\omega_{\text{gate}})$ should have the same OAM as the signal field $E_l(\mathbf{r}, \omega_{\text{signal}})$.

Figure \ref{fig:oam_images} compares the spatial profiles of the signal field with those of the sum-FWM field without and with the 600-grit and 1500-grit ground glass diffusers combined. From Fig. \ref{fig:oam_images}B, we note that scattering due to diffusers adds noise to the spatial profile of the signal, but its donut shape remains preserved, along with its topological charge, as shown by the cylindrical lens transform (Figs. \ref{fig:oam_images}C and \ref{fig:oam_images}D). However, for the sum-FWM field, both the transverse spatial profile (Figs. \ref{fig:oam_images}E and \ref{fig:oam_images}F) and the topological charge (Figs. \ref{fig:oam_images}G and \ref{fig:oam_images}H) show robustness to scattering. Table \ref{tab:psn_oam} shows the PSNR of the scattered signal and the sum-FWM for all the diffusers considered. See Fig.~\ref{fig:oam_scattering} in the supplementary materials for the corresponding images of the scattered signal field and the gated sum-FWM field. As before, we calculate the PSNR from the spatial profiles of the scattered signal field and the sum-FWM field with their respective profiles in the absence of scattering used as references. As the scattering strength increases, the PSNR of the scattered signal field drops significantly from 20.31 dB to 15.27 dB. On the other hand, the PSNR of the sum-FWM field remains stable around 20 dB. Specifically for the 600-grit diffuser and the combined diffuser stack (600-grit plus 1500-grit), a 4-dB difference in PSNR between the scattered signal field and the sum-FWM field shows an excellent rejection of scattering by the time-gate.

\subsection*{Real-time removal of scattering}
Real scattering media, such as a cloud cover or biological tissues, are dynamic, as the light-scattering particles are in constant random motion. Therefore, the real-time removal of scattering and scintillation is an important factor to consider in the design of any imaging or optical communication modality in such systems. In this section, we show that our time-gate is also highly effective in removing the scintillation from dynamic or time-varying scattering media. 

In the first demonstration, we use polystyrene beads of average diameters of 7.7 $\mu$m suspended in an aqueous solution as the dynamic scattering medium. This suspension is contained in a 2-mm-thick glass cuvette and placed after the object in the path of the signal pulse. The bead suspension functions as an excellent time-varying scattering medium because of the constant random motion of the beads in water and is often used in biomedical imaging to mimic light scattering in biological tissues. Table \ref{tab:psnr_pcc_latex} compares the image quality metrics (PSNR and PCC) of the freeze frames of the scattered signal with the gated sum-FWM image at various bead concentrations. Similar to the results in Table ~\ref{tab:psnr_pcc_images}, both metrics of the sum-FWM images remain significantly higher than the direct images for all bead concentrations considered. The freeze frames of the scattered signal, and the gated sum-FWM image at various bead concentrations are shown in Fig. \ref{fig:latex_beads} in the supplementary materials.

To quantify the performance of our time-gate for dynamic scattering media, we monitor the intensity fluctuations of the scattered signal and gated sum-FWM images over a fixed period of time. Supplementary movies S1 and S2 clearly show that the image quality of the sum-FWM field remains stable within a 10-second time interval, whereas the signal field shows a strong scintillation effect as a result of the random motion of beads. To quantify these intensity fluctuations, or ``scintillation", we define the scintillation index $\sigma_s$ at each pixel  as
\begin{align}\label{eq:scint_index}
    \sigma_s = \frac{\left<I^2\right>}{\left<I\right>^2} - 1,
\end{align} where $\left<\cdot\right>$ denotes the time average. We uniformly split the 10-second video into 200 frames and calculate the scintillation index for each pixel in similar regions of interest for both signal and sum-FWM images. The resultant scintillation index map thus indicates the stability of the image in a dynamic scattering medium. 

Figures \ref{fig:latex_beads_image}A and Fig. \ref{fig:latex_beads_image}B show the scintillation maps of the direct images of the test object (Fig.~\ref{fig:info_images}A) in the path of the signal pulse, and of the scattering-free upconverted sum-FWM images at a bead concentration of 0.525 g/cm$^3$. We note that the scintillation map of the signal images (Fig. \ref{fig:latex_beads_image}A) shows significant pixel-to-pixel variation in $\sigma_s$ in contrast to the scintillation index map for the sum-FWM images (Fig. \ref{fig:latex_beads_image}B), which has almost negligibly small values of $\sigma_s$ throughout. Hence, the scattered signal field has large intensity fluctuations over time, whereas the gated sum-FWM field is essentially a static image with all the artefacts due to scattering removed. As such, the histograms of $\sigma_s$ for the signal field (Figs. \ref{fig:latex_beads_image}C and Fig. \ref{fig:latex_beads_image}D) show a much wider distribution than for the corresponding sum-FWM field (Figs. \ref{fig:latex_beads_image}G and Fig. \ref{fig:latex_beads_image}H) with values that are approximately two orders of magnitude larger.

In the second demonstration, we slowly rotate the ground glass diffusers and monitor the scintillation in the spatial profile of the OAM-encoded signal field. Supplementary movie S3 shows the intensity profiles of the scattered signal field and the gated sum-FWM field as the combined diffuser stack (1500-grit and 600-grit) is rotated, and Figs. \ref{fig:latex_beads_image}E-H the respective scintillation maps and the associated histogram counts. We measure an order of magnitude larger overall values of $\sigma_s$ with a wider distribution in this instance compared to the latex beads for the signal field. In contrast, we measure negligible scintillation for the gated sum-FWM field. In both studies, the larger values of $\sigma_s$ are obviously concentrated in the regions of intensity minima of the original fields as $\sigma_s$ is normalized to the time-averaged intensity, which is homogenized over the fluctuations and approaches the original unscattered intensity distributions.

\subsection*{Discussion and conclusion}
The sum-FWM process in our implementation of the time gate upconverts the frequency of the image-carrying beam into the visible region, thereby enabling easier photon detection through efficient silicon-based detectors over InGaAs \cite{barh2019parametric}. We note that phase-matching is difficult to achieve for sum-FWM in most bulk media due to dispersion and is not an efficient process. Sum-FWM has previously been reported in systems with a very short interaction length, such as dielectric metasurfaces \cite{liu2018all}, highly focused beams in gases and atomic vapor \cite{wallace1976high, hilber1987broadly, hodgson1974tunable}, and in an ENZ film \cite{jaffray2022near}. The tunability of the ENZ wavelength of TCOs through doping, and by integrating with metasurfaces \cite{alam2018large}, also provides tremendous flexibility to customize the gate for the wavelength of choice. Furthermore, our gating technique could be extended to materials that have an ENZ wavelength at mid-IR frequencies such as doped semiconductors like CdO \cite{runnerstrom2018polaritonic} and polar dielectrics \cite{passler2018strong}, which would enable the upconversion of ballistic mid-IR photons to recover images at visible or NIR wavelengths. The mid-IR imaging and sensing through scattering media is essential in chemical sensing \cite{christ2015chemical}, non-destructive biomedical imaging \cite{zhang2022near}, and detection of explosives \cite{ma2014upconversion} but is particularly limited by the lack of sensitive and fast detectors.

Furthermore, we compare the performance of our time gate with two other mainstream time-gating methods in Table \ref{tab:compare}. The traditional Kerr gate uses a CS$_2$ cell as the nonlinear medium \cite{wang1991ballistic, wang1993time, alfano1994time}, which in addition to the instantaneous electronic nonlinearity \cite{yan2008influence}, also has a slower nonlinearity due to molecular orientation \cite{boyd2020nonlinear}. Consequently, the actual width of the time gate can reach more than 500 fs (FWHM) for a 100-fs incident gating pulse \cite{yan2008influence, tong2012femtosecond}, which would lead to a suboptimal scattering rejection. Other time gates based on second- or third-order nonlinear processes in bulk crystals such as sum/difference frequency generation (SFG/DFG) \cite{yoo1991imaging, watson1995imaging, andreoni2000viewing, faris1994upconverting} and difference-FWM \cite{sappey1994optical} have also been shown to increase the signal-to-noise ratio and the brightness of the signal. These modalities based on fast electronic responses do reduce the width of the gate to that of the incident pulse and also enable frequency conversion. However, the tight focusing of the image-carrying beam and the gate beam inside the bulk crystals leads to a loss of resolution because of the limits set forth by the phase-matching condition, the small spot size of the focused gate beam and the limited acceptance angle of the nonlinear crystal \cite{faris1994upconverting}. In this context, our sub-wavelength-thick ENZ film with a large nonlinearity relaxes the phase-matching condition and thus generates an efficient gated signal with a complete rejection of scattering and no loss of spatial resolution. Finally, the recent advancement in light-guiding through complex media using computational wavefront shaping methods \cite{mididoddi2025threading, haim2025image} have shown great adaptability to various scattering conditions. We expect that our all-optical time-gate could alleviate the computational overhead of the aforementioned methods by pre-selecting the quasi-ballistic photons. 

In conclusion, we have demonstrated the first transfer of spatial information encoded in both intensity and phase through scattering media along with upconversion using the enhanced instantaneous third-order nonlinear response of ITO in the ENZ regime. The sub-picosecond nonlinear response and the large field enhancement in ENZ materials make them an excellent platform for transmitting signals through static or time-varying scattering media. Furthermore, the efficient upconversion FWM process in our gate enables signal recovery at visible wavelengths without compromising the imaging resolution or the rapid temporal response. Finally, we have studied the robustness of our time gate with respect to time-varying scattering media and have shown an excellent rejection of scintillation, thereby enabling real-time retrieval of signal with high fidelity. This dynamic gating of ballistic photons is highly relevant for both underwater and free-space optical communication, and for non-invasive \textit{in vivo} biomedical imaging. 

\newpage

\begin{figure}
  \centering
  \includegraphics[width=\textwidth]{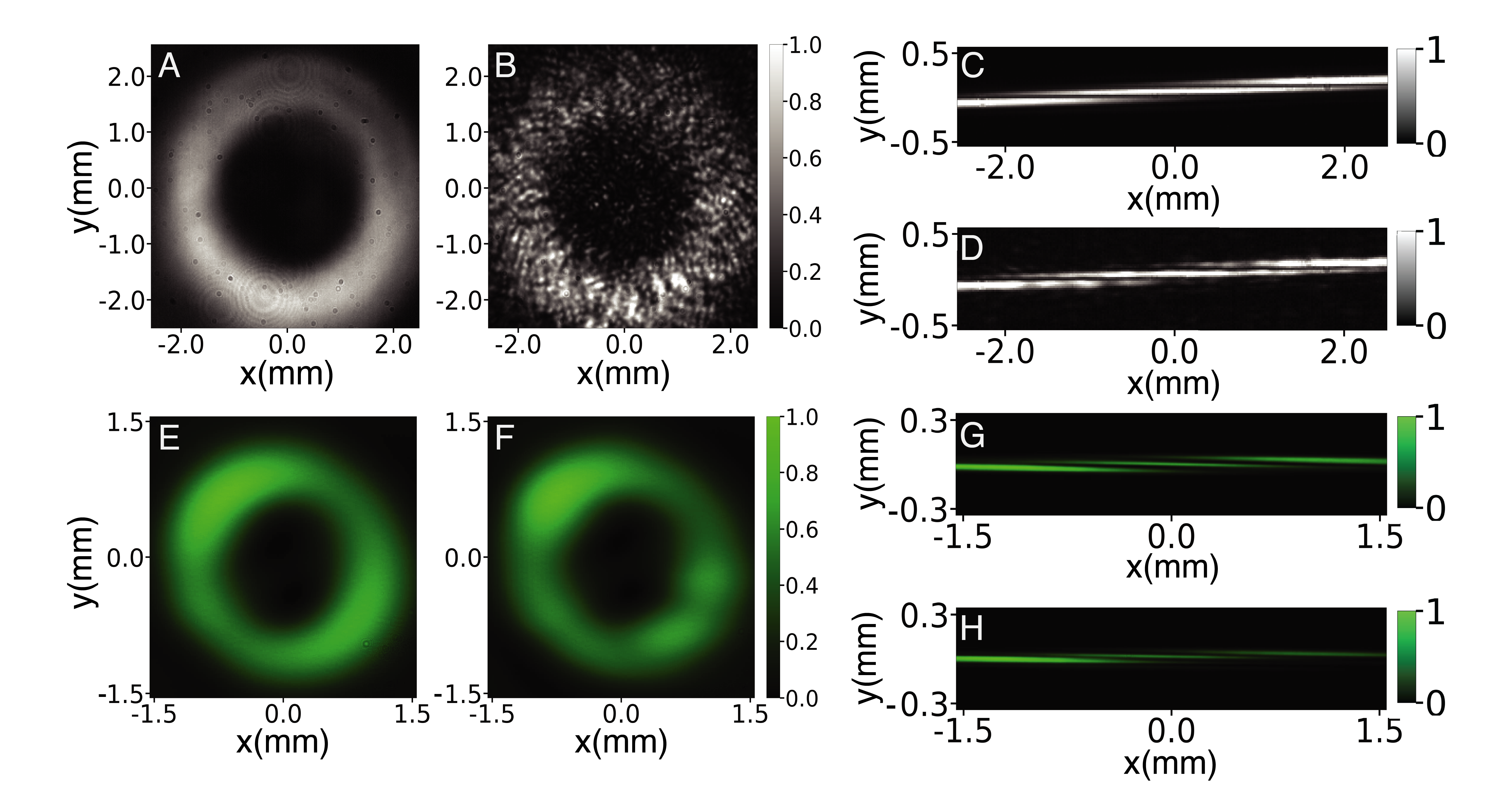}
\caption{\textbf{The transfer of amplitude and phase distributions of OAM-carrying pulses through ground glass diffusers with a sum-FWM time-gate.} The spatial profiles of the \textbf{(A)} incident signal field and the \textbf{(E)} gated sum-FWM field with no scatterers. Both the signal and the sum-FWM fields carry an OAM of topological charge $l = 2$, which is evident in their spatial profiles at the focus of a cylindrical lens (or the cylindrical lens transform) shown in panels \textbf{(C)} and \textbf{(G)}, respectively. The spatial profile of \textbf{(B)} the signal field with the combined (1500- and 600-grit) diffusers shows significant noise in the intensity distribution, whereas the spatial profile of \textbf{(E)} the sum-FWM field is relatively unaffected by scattering. \textbf{(D)} The cylindrical lens transform of the scattered signal field shows that the topological charge is largely preserved despite the scattering. \textbf{(H)} The cylindrical lens transform of  gated sum-FWM field with the diffusers in place shows negligible difference with the unscattered counterpart in panel \textbf{(G)}.}
\label{fig:oam_images}
\end{figure}

\begin{figure}
  \centering
  \includegraphics[width=0.55\textwidth]{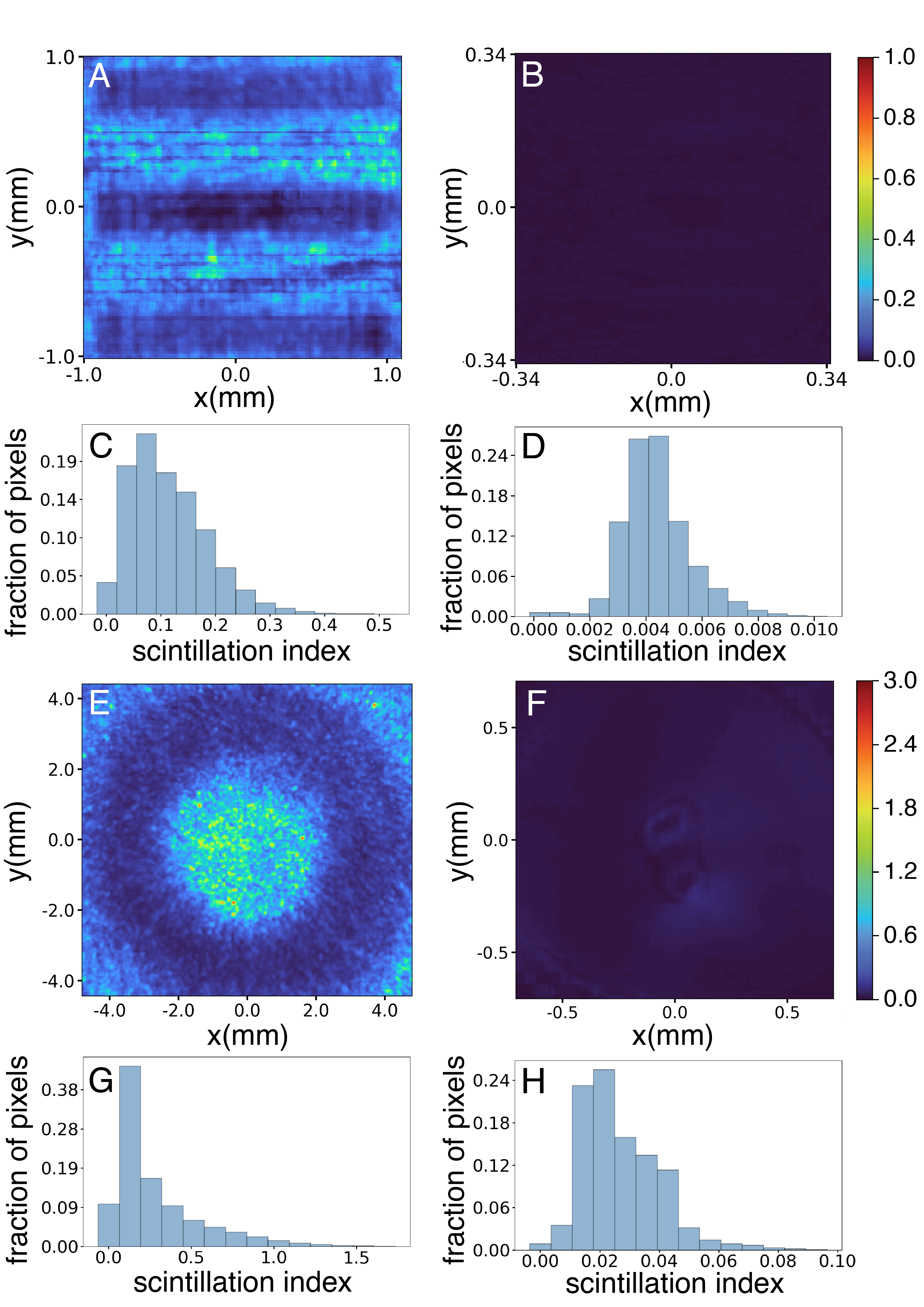}
\caption{\textbf{The suppression of scintillation due to dynamic scattering.} Scintillation index maps of the USAF target (shown in Fig. \ref{fig:info_images}A) in the \textbf{(A)} direct image of the scattered signal, and in the \textbf{(B)} gated sum-FWM image. The associated normalized histogram counts are shown in \textbf{(C)} and \textbf{(D)}. The scattering medium is the aqueous suspension of moving latex beads with a mass concentration of 0.525 g/cm$^3$ (see Fig. \ref{fig:latex_beads}A, B in the supplementary materials for the results at lower concentrations). The scintillation effect due to dynamic light scattering is suppressed by two orders of magnitude in the sum-FWM image. \textbf{(E)-(H)} Scintillation index maps and the associated histogram counts of the OAM-encoded signal field and the sum-FWM field shown in Fig. \ref{fig:oam_images}. The dynamic scatterer is created by rotating the combined ground glass diffusers (1500-grit and 600-grit) (see Fig. \ref{fig:oam_scattering}C, D in the supplementary materials for the results with the individual diffusers). The gated sum-FWM field shows minimal intensity fluctuations despite a strong scattering of the incident signal field. The scintillation index of each pixel is computed through the use of Eq. \ref{eq:scint_index} from 200 frames taken within 10 seconds. }
\label{fig:latex_beads_image}
\end{figure}

\newpage

\begin{table}
\caption{\textbf{Peak signal-to-noise ratio (PSNR) and Pearson correlation coefficient (PCC) of the scattered signal images and the sum-FWM images for different ground glass diffusers. }}\label{tab:psnr_pcc_images}
\begin{tabular*}{\textwidth}{@{\extracolsep\fill}lcccccc}
\\
\hline
& \multicolumn{2}{@{}c@{}}{Scattered Signal at 1550 nm} & \multicolumn{2}{@{}c@{}}{Sum-FWM at 517 nm} \\%
Scattering Medium & PSNR (dB) & PCC &  PSNR (dB) & PCC\\
\hline
1500-grit ground glass& 16.25 & 0.887 & 24.61 & 0.940\\
600-grit ground glass & 10.54 & 0.373 & 27.77 & 0.962\\
600-grit \& 1500-grit combination & 9.78  & 0.219  &  24.77 & 0.940\\
\hline
\end{tabular*}
\end{table}

\begin{table}
\caption{\textbf{Peak signal-to-noise ratio (PSNR) of the transverse spatial profile of the signal field carrying $l=2$ and the corresponding sum-FWM field.}}\label{tab1}%
\begin{tabular}{@{}llll@{}}
\\
\hline
Scattering Medium & Signal PSNR (dB)  & Sum-FWM PSNR (dB)\\
\hline
1500-grit  & 20.31 & 20.64 \\
600-grit  & 15.87  & 19.74 \\
\text{600-grit \& 1500-grit combination}  & 15.27  & 19.27 \\
\hline
\end{tabular}
\label{tab:psn_oam}
\end{table}

\begin{table}
\caption{\textbf{Peak signal-to-noise ratio (PSNR) and Pearson correlation coefficient (PCC) of the direct images and the sum-FWM images of the USAF target for latex beads suspension with different mass concentrations.} }\label{tab:psnr_pcc_latex}
\begin{tabular*}{\textwidth}{@{\extracolsep\fill}lcccccc}
\\
\hline
& \multicolumn{2}{@{}c@{}}{Scattered Signal at 1550 nm} & \multicolumn{2}{@{}c@{}}{Sum-FWM at 517 nm} \\
Scattering Medium & PSNR (dB) & PCC &  PSNR (dB) & PCC\\
\hline
0.175 g/cm$^3$ latex beads suspension & 15.85  & 0.878  &  20.36 & 0.976\\
0.350 g/cm$^3$ latex beads suspension & 15.28  & 0.856  &  21.64 & 0.920\\
0.525 g/cm$^3$ latex beads suspension & 15.24  & 0.850  &  20.78 & 0.961\\
\hline
\end{tabular*}
\end{table}

\begin{table}
\caption{\textbf{Comparison of our ENZ-based sum-FWM gate with other time-gating methods for imaging through scattering media.}}\label{tab:compare}%
\begin{tabular}{@{}llll@{}}
\\
\hline
Imaging Methods & Gate Width  & Resolution Loss & Wavelength Conversion\\
\hline
CS$_2$ Kerr gate& longer than gating pulse  & Yes  & Not possible \\
Upconversion gate& pulse-width limited & Yes & Possible \\
Sum-FWM on ENZ & pulse-width limited & No  & Possible \\
\hline
\end{tabular}
\end{table}


\clearpage 

%
\bibliography{main} 
\bibliographystyle{sciencemag}

%
%
%
%
%
%


\section*{Acknowledgments}
The authors thank Edouard Berrocal at Lund University for his generous help in Monte Carlo simulation of light scattering in turbid media. The authors would also like to thank J.W. Kuper for useful discussions.
\paragraph*{Funding:}
This work was supported by the US Office of Naval Research through MURI award N00014-20-1-2558 and through award N00014-19-1-2247 and by the US National Science Foundation under Award 2138174.  In addition, M.Z.A. and R.W.B. acknowledge support by  the Natural Sciences and Engineering Research Council of Canada under the Alliance Consortia Quantum Grant ALLRP 578468 - 22 and Discovery Grant RGPIN/2017-06880 and by the Canada Research Chairs program under award 950-231657.

\paragraph*{Author contributions:}
Y.X., S.C., M.Z.A. and R.W.B. initiated the project. L.D.N. produced the imaging targets and assisted in the experiment setup. Y.X. and S.C. carried out the time-gated imaging and OAM transfer measurement. M.K., S.V. and J.R.H. prepared the ITO samples and performed the ENZ characterization measurements. Y.X. carried out the scattering simulation. J.K.M. and E.G.J. fabricated the phase plate for OAM generation. Y.X. and S.C. analyzed the data. Y.X., S.C., M.Z.A and R.W.B wrote the paper with input from all authors.

\paragraph*{Competing interests:}
There are no competing interests to declare.

\paragraph*{Data and materials availability:}
All data needed to evaluate the conclusions in the paper are present in the paper or supplementary materials.


\subsection*{Supplementary materials}
Materials and Methods\\
Supplementary Text\\
Figs. S1 to S10\\
References \textit{(81-\arabic{enumiv})}\\ 
Movie S1 to S3\\


\newpage


\renewcommand{\thefigure}{S\arabic{figure}}
\renewcommand{\thetable}{S\arabic{table}}
\renewcommand{\theequation}{S\arabic{equation}}
\renewcommand{\thepage}{S\arabic{page}}
\setcounter{figure}{0}
\setcounter{table}{0}
\setcounter{equation}{0}
\setcounter{page}{1} 


\begin{center}
\section*{Supplementary Materials for\\ \scititle}

Yang~Xu$^{1\ast\dagger}$,
Saumya~Choudhary$^{2\dagger}$,
Long~D.~Nguyen$^{1}$,
Matthew~Klein$^{3,4}$,\\
Shivashankar Vangala$^{4}$, J. Keith Miller$^{5}$,
Eric G. Johnson$^{6}$, \\
Joshua R. Hendrickson$^{4}$, M. Zahirul Alam$^{7}$,Robert~W.~Boyd$^{1,2,7}$\\

\small$^\ast$Corresponding author. Email: yxu100@ur.rochester.edu\\
\small$^\dagger$These authors contributed equally to this work.
\end{center}

\subsubsection*{This PDF file includes:}
Materials and Methods\\
Supplementary Text\\
Figures S1 to S10\\
Captions for Movies S1 to S3\\

\subsubsection*{Other Supplementary Materials for this manuscript:}
Movies S1 to S3\\

\newpage


\subsection*{Materials and Methods}
\subsubsection*{Experiment Setup}
\begin{figure}
  \centering
  \includegraphics[width=\textwidth]{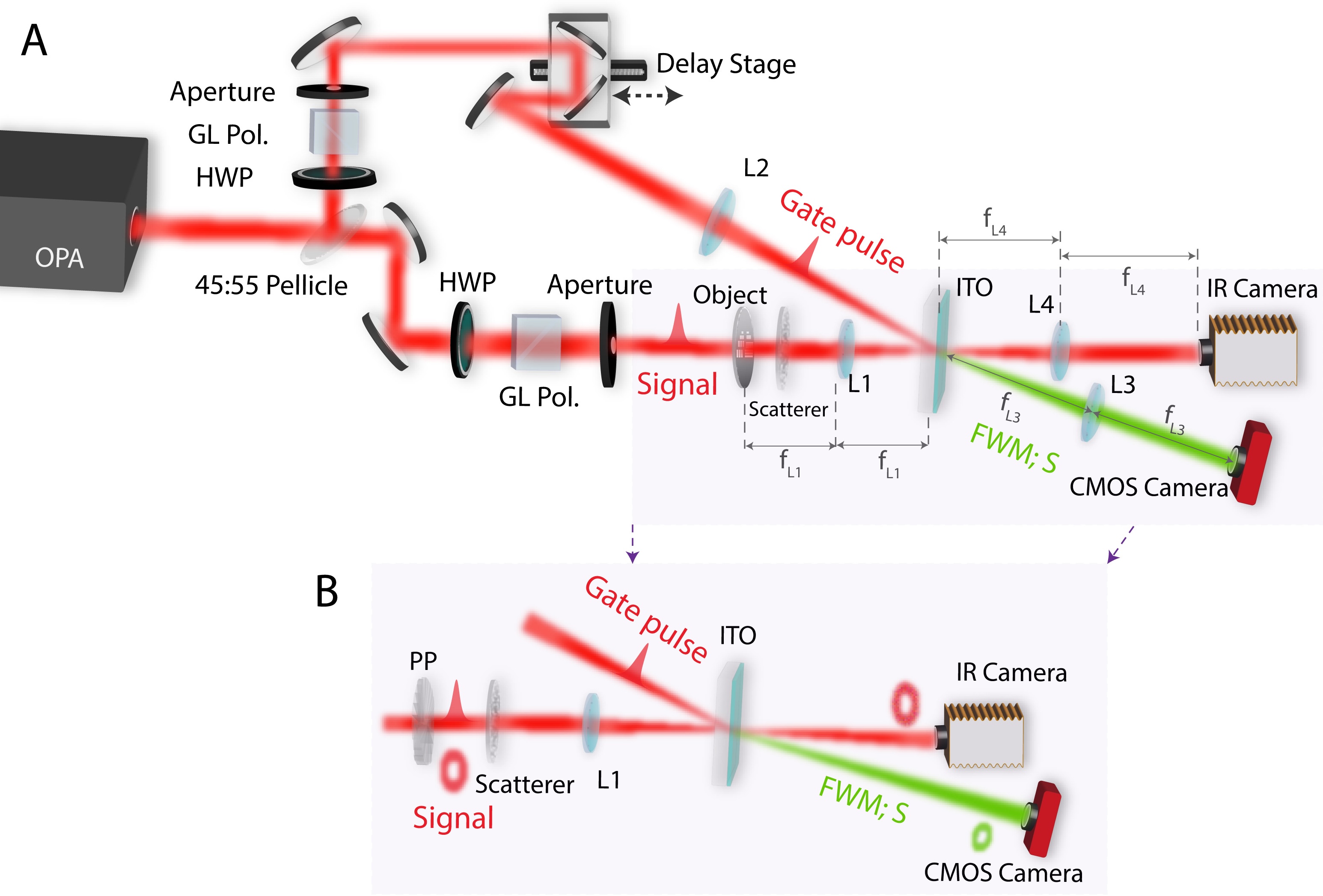}
\caption{\textbf{The experimental setup.} A schematic of our experimental setup for the sum-FWM time-gate for \textbf{(A)} imaging and \textbf{(B)} transfer of OAM through scattering media. HWP: half-wave plate; GL Pol: Glan-Taylor polarizer; PP: spiral phase plate; L: plano-convex lenses.}

\label{fig:setup}
\end{figure}

The schematic of our setup for the sum-FWM time-gate is shown in Fig. \ref{fig:setup}.  Figure \ref{fig:setup}A shows the configuration for recovering images, and Fig. \ref{fig:setup}B shows the configuration for the transfer of OAM. In both setups, the OPA (Coherent OPerA Solo) produces our 100 fs-wide source pulse centered at 1550 nm with a repetition rate of 1 kHz. The source pulse is split evenly into gate and signal pulses. The signal beam then passes through the (a) object (spiral phase plate) followed by the scattering medium and is then focused by a plano-convex lens ($f = 20$ mm) onto the ITO sample from the side of the substrate. The gate pulse goes through a delay stage, which controls the relative optical path delay between the gate and the signal pulses, and is gently focused onto the thin-film ITO by a lens with a longer focal length ($f = 300$ mm). The different focusing lenses for the gate pulse and signal pulse yield a focal spot of the gate beam that is approximately 7$\times$ larger than the signal, which is required to have efficient nonlinear conversion from the tails of the angular spectrum of the signal when the object is in place to preserve the imaging resolution. The sum-FWM is collected on a CMOS camera (Thorlabs CS165MU) with or without an imaging lens for configurations A and B, respectively. The signal is also similarly collected on a stabilized NIR camera (Xenics Bobcat 640) with or without an imaging lens. The intensities of the gate and the signal fields without any scatterers were 70.5 GW/cm$^{2}$ and 184.1 GW/cm$^{2}$, respectively. We adjust the temporal delay between the gate pulse and the scattered signal pulse to overlap with the early-arriving ballistic photons. The generated sum-FWM field is completely free from any scattering and is also bright enough to be collected by the CMOS camera. We place the scatterers right after the object or the phase plate in configurations A and B, respectively. 

\subsubsection*{Calculation of the peak signal-to-noise ratio (PSNR) and the Pearson correlation coefficient (PCC) }
Peak signal-to-noise ratio (PSNR) is a commonly used metric for image quality, and is given by the ratio of the maximum power of the signal and the power of the distorting noise. The PSNR of a given image can be calculated from the following expression 
\begin{align}
    \text{PSNR} = 20 \, \text{log}_{10} \left(\frac{R}{\sqrt{\text{MSE}}}\right) 
\end{align}
where $R$ is the maximum signal power in the reference image and MSE is the mean squared error defined as
\begin{align}
    \text{MSE} = \frac{1}{MN}\sum_{i=1}^M\sum_{j=1}^N \left|x_{ij}-y_{ij}\right|^2. 
\end{align}
Here, $x_{ij}$ ($y_{ij}$) stands for the gray-scale value of the pixel at the $i^{\text{th}}$ row and $j^{\text{th}}$ column of the $M \times N$ tested (reference) image. 

The Pearson correlation coefficient (PCC) is also a computational metric for describing the similarity between two images and is given by the linear correlation between the pixel arrays of the two images. PCC is mostly used in image processing for pattern recognition and image classification, and can be calculated from the following expression
\begin{align}
    \text{PCC} = \frac{\sum_i (x_i - x_m)(y_i - y_m)}{\sqrt{\sum_i (x_i-x_m)^2}\sqrt{\sum_i (y_i-y_m)^2}}.
\end{align}
Here $x_i$ ($y_i$) is the intensity of the $i^{\text{th}}$ pixel in the first (second) image, and $x_m$ ($y_m$) is the mean intensity of all pixels in the first (second) image. The value of PCC ranges from $-1$ to 1. A PCC of 1 ($-1$) shows that the pixels in the two images are completely correlated (anti-correlated). In our calculations, all images are read in 8-bit gray-level format and centered and cropped to the same size prior to computing both metrics.


\subsection*{Supplementary Text}
\subsubsection*{The effect of scattering on the spatial and temporal profiles of photons}
We use the open-source software, \textit{Multi-Scattering}, to simulate the effect of scattering in an imaging system with a negative USAF-1951 target as the object \cite{frantz2022multi, jonsson2020multi}. The software is based on the Monte-Carlo simulation that tracks millions of individual photons that propagate through cubic scattering volumes. \textit{Multi-Scattering} uses the Lorenz-Mie theory to generate the phase function from spherical particles to model the scattering of light. 
    
\begin{figure}
    \centering
    \includegraphics[width=0.75\textwidth]{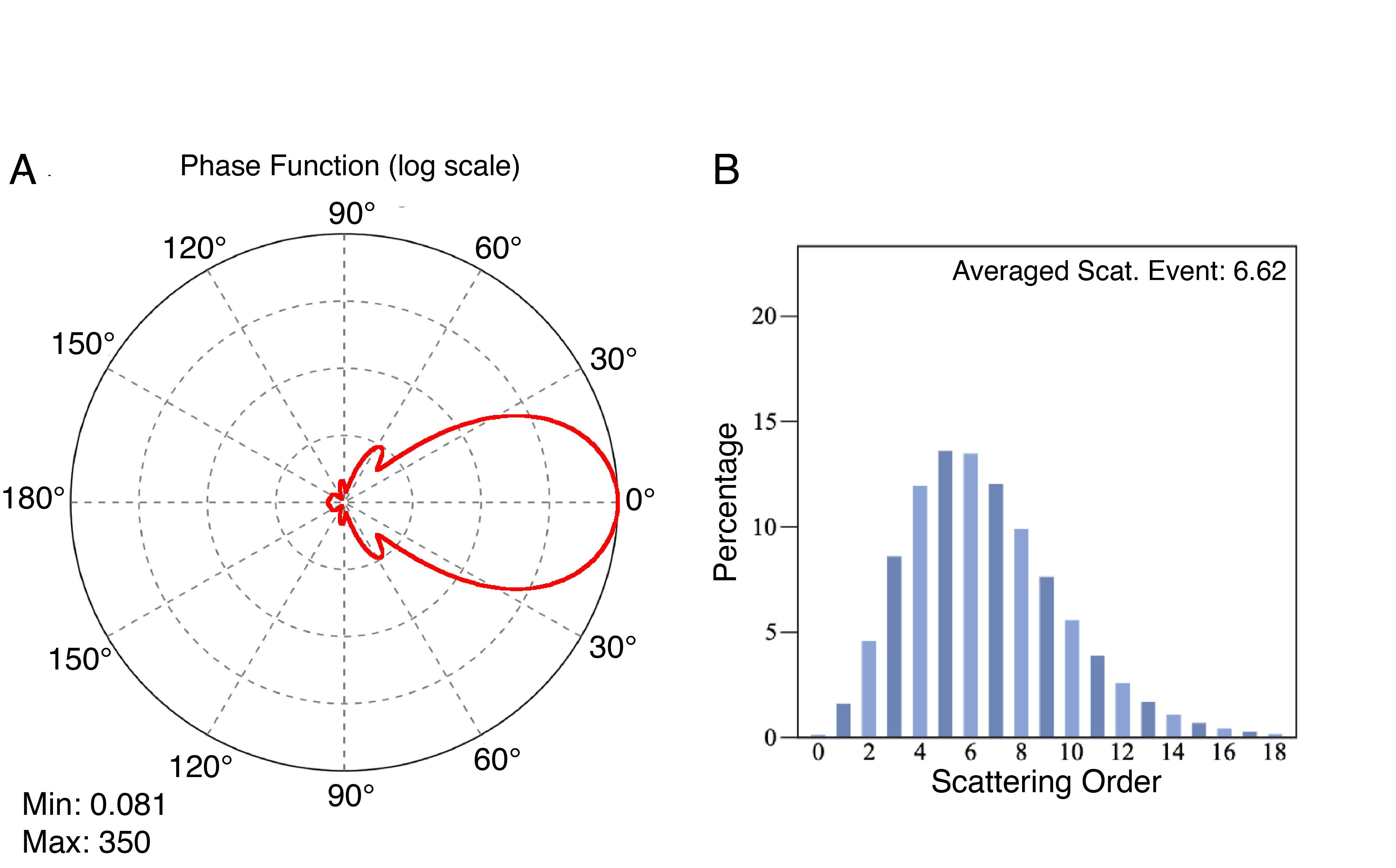}
    \caption{\textbf{Metrics of the scatterer considered for the Mie scattering simulations.} \textbf{(A)} The phase function of the aqueous suspension of polystyrene beads of average diameters of 7.7 $\mu$m at a concentration of 0.525 g/cm$^3$. The phase function characterizes the angular distribution of all scattered photons after the 2-mm-thick cuvette holding the bead suspension. The computed phase function shows that the scattering due to the beads is predominantly in the forward direction. \textbf{(B)} The scattering order distribution of the scattered photons, which indicates the total number of scattering events that a photon undergoes when it propagates through the medium. For the bead suspension described in \textbf{(A)}, the incident photons are scattered 6.62 times on average. We note that the ballistic (scattering order = 0) and quasi-ballistic (scattering order = 1) photons comprise a very small fraction of the scattered photons. }
    \label{fig:scatterer_info}
    \end{figure}

For a fair comparison with our experimental results shown in Fig. 3 in the main text, we specify the scattering medium in our simulation to be a 2 mm $\times$ 2 mm $\times$ 2 mm cube filled with an aqueous suspension of polystyrene beads. The concentration of the suspension is held at 0.525 $\text{g}/\text{cm}^3$, and the beads have an average diameter of 7.7 $\mu$m with a refractive index at room temperature described by the Sellmeier equation \cite{sultanova2009dispersion} as 
    \begin{align}
        n^2 - 1 = \frac{1.4425\lambda^2}{\lambda^2 - 0.020216}\label{eq:ref_index},
    \end{align}
where $\lambda$ is the wavelength of the incident light in microns. The central wavelength of the illuminating pulse in our experiment is 1.55 $\mu$m. From the physical parameters of the scattering medium described above, we can generate the phase function (Fig. \ref{fig:scatterer_info}A) and the scattering order distribution (Fig. \ref{fig:scatterer_info}B) for the bead suspension. 

\begin{figure}
    \centering
    \includegraphics[width=\textwidth]{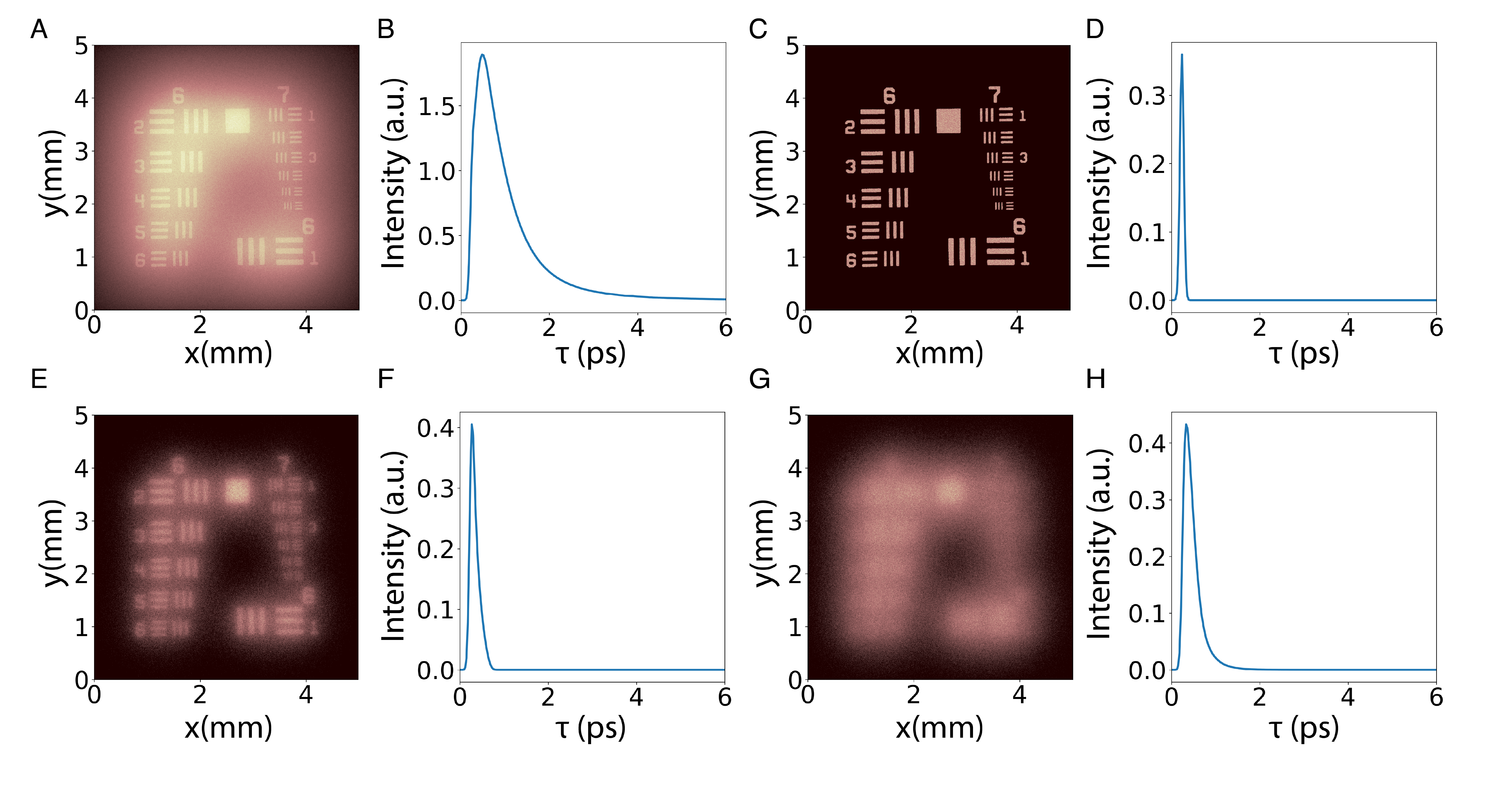}
    \caption{\textbf{Spatial and temporal profiles of ballistic and diffusive photons after scattering.} The spatial (density plots) and temporal (line plots) profiles of \textbf{(A), (B)} all forward-scattered photons, \textbf{(C), (D)} all ballistic photons, \textbf{(E), (F)} the photons scattered only once (quasi-ballistic), and \textbf{(G), (H)} the photons scattered 6 times inside the medium.}
    \label{fig:ballistic_image}
    \end{figure}

Figure \ref{fig:ballistic_image} shows the spatial and temporal profiles of the photons collected at the image plane with different scattering orders. The illuminating pulse used in our simulation has a Gaussian temporal profile with a full-width-half-maximum (FWHM) of 100 fs and an average photon number of $10^9$ per pulse. The USAF-1951 target is imaged with a single-lens 2f-2f system. Figures \ref{fig:ballistic_image}A and \ref{fig:ballistic_image}B show the spatial profile and the temporal profile of all the scattered photons arriving at the image plane, respectively. We notice a decrease in the image quality and a significant increase in the overall pulse width due to scattering. Figure \ref{fig:ballistic_image}C shows the image formed by just the ballistic photons, and is essentially a perfect image of the  test object. Their significantly narrower temporal profile in Fig. \ref{fig:ballistic_image}D compared to Fig. \ref{fig:ballistic_image}B confirms that the ballistic photons travel through the medium without being scattered. Figures \ref{fig:ballistic_image}E-F show the spatial and temporal profiles of the quasi-ballistic photons, and Figs. \ref{fig:ballistic_image}G-H the profiles of the photons scattered 6 times inside the medium. We note that the image becomes more obscured and the pulse width becomes longer as more scattered photons are used to form the image. 

\subsubsection*{The permittivity spectrum and the nonlinear response of ITO}
The large nonlinearity of an epsilon-near-zero (ENZ) material such as ITO has two different origins: an instantaneous electronic nonlinear response and a slower response due to the thermal excitation of electrons \cite{khurgin2021fast}. The instantaneous nonlinearity is often treated as the perturbation of the electron motion in a non-quadratic potential \cite{boydnlo}. Some well-studied examples of this fast nonlinearity include wave mixing processes and harmonics generation. On the other hand, the slow nonlinearity is often associated with real excitations as opposed to virtual transitions that lead to instantaneous nonlinearity. The signature of the slow component is its longer lifetime which usually results from the relaxation of heated electrons. A typical example is the large nonlinear refractive index change of ITO at the ENZ wavelength.

Because of the unprecedentedly large nonlinear response of ITO, the forward four-wave mixing (FFWM) process shows features not usually observed in FFWM. In particular, as illustrated in Fig.~\ref{fig:fwm_ito}B, four FFWM processes are observed, two each for sum-frequency mixing and for difference frequency mixing, giving rise to output frequencies $\omega_{\rm FWM, \, S} = 2 \omega_{\rm gate} + \omega_{\rm signal}$, $\omega_{\rm FWM, \, S^*} = \omega_{\rm gate} + 2 \omega_{\rm signal}$, $\omega_{\rm FWM, \, D} = 2 \omega_{\rm gate} - \omega_{\rm signal}$, and $\omega_{\rm FWM, \, D^*} = 2\omega_{\rm signal} - \omega_{\rm gate}$. Here, the unstarred versions involve two photons from the gate field and one photon from the signal field; the starred versions involve two photons from the signal field and one gate photon. For a weak signal field, the starred version would be expected to be much weaker than the unstarred version. It is only because of the large nonlinear response of ITO that the unstarred versions become observable. Another consequence of the large nonlinear response is that phase-matching conditions do not limit the processes that can occur. Because the nonlinear response is very large, the nonlinear material can be very thin (only a few hundred nm in our work), and thus the phase mismatch factor $\Delta k L$ is usually much smaller than unity for any value of the longitudinal wavevector mismatch $\Delta k$. Thus, while the transverse components of the wavevectors need to be conserved, the longitudinal components do not. 

\subsubsection*{The time scales of the two types of nonlinear responses at ENZ}
\begin{figure}
    \centering
    \includegraphics[width=0.5\textwidth]{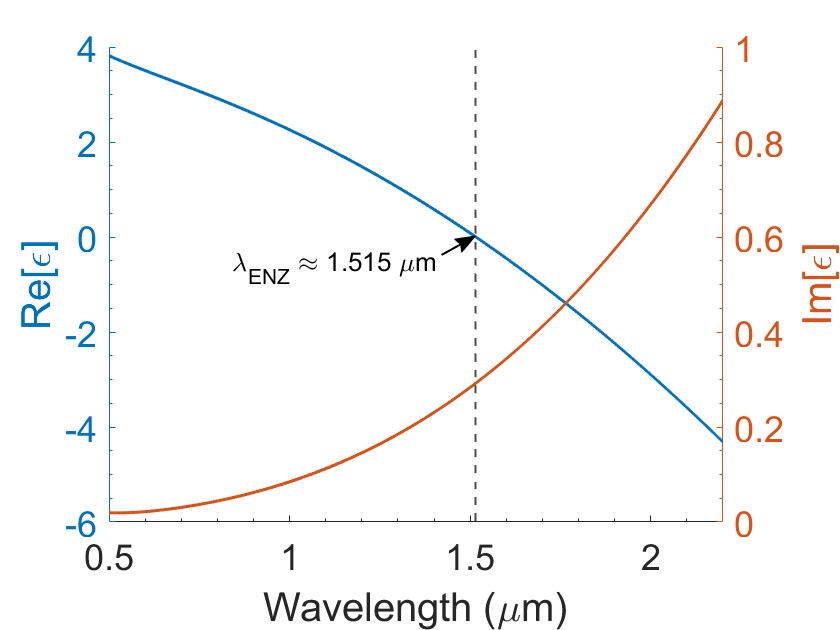}
    \caption{\textbf{The epsilon-near-zero (ENZ) region of the ITO used.} The spectra of real and imaginary parts of the permittivity of ITO sample in our experiment. The real part of the permittivity crosses zero around a wavelength of 1.515 $\mu$m.}
    \label{fig:ito_epsilon}
    \end{figure}

\begin{figure}
    \centering
    \includegraphics[width=\textwidth]{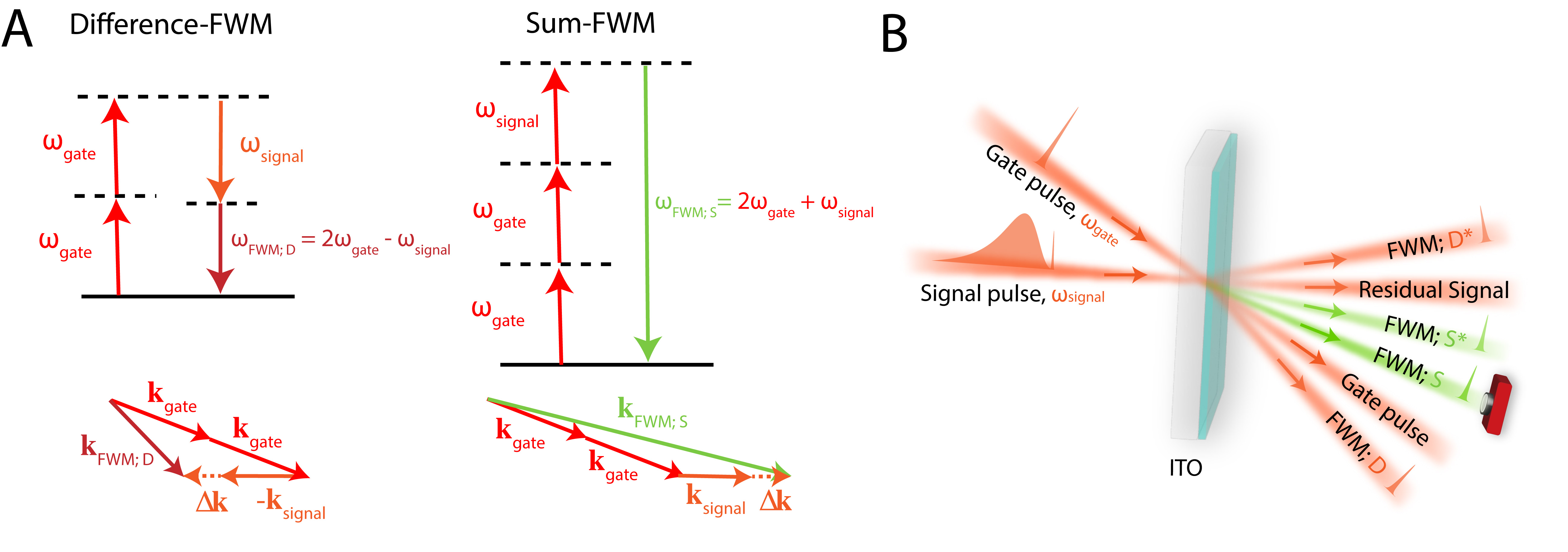}
    \caption{\textbf{The energy and momentum conservation relations of sum- and difference-FWM in ITO.} \textbf{(A)} The energy level diagrams and momentum conservation relations for the difference and sum-FWM processes that occur in ITO between the gate and signal pulses in the configuration shown in \textbf{(B)}. The difference-FWM pulses (D and D*) emerge at larger angles than both the exciting pulses and the sum-FWM pulses. The D- and S-FWM pulses are produced by the mixing of two gate photons and one signal photon, whereas the D* and S* pulses are produced by the mixing of two signal photons and one gate photon.}
    \label{fig:fwm_ito}
    \end{figure}

\begin{figure}
    \centering
    \includegraphics[width=0.9\textwidth]{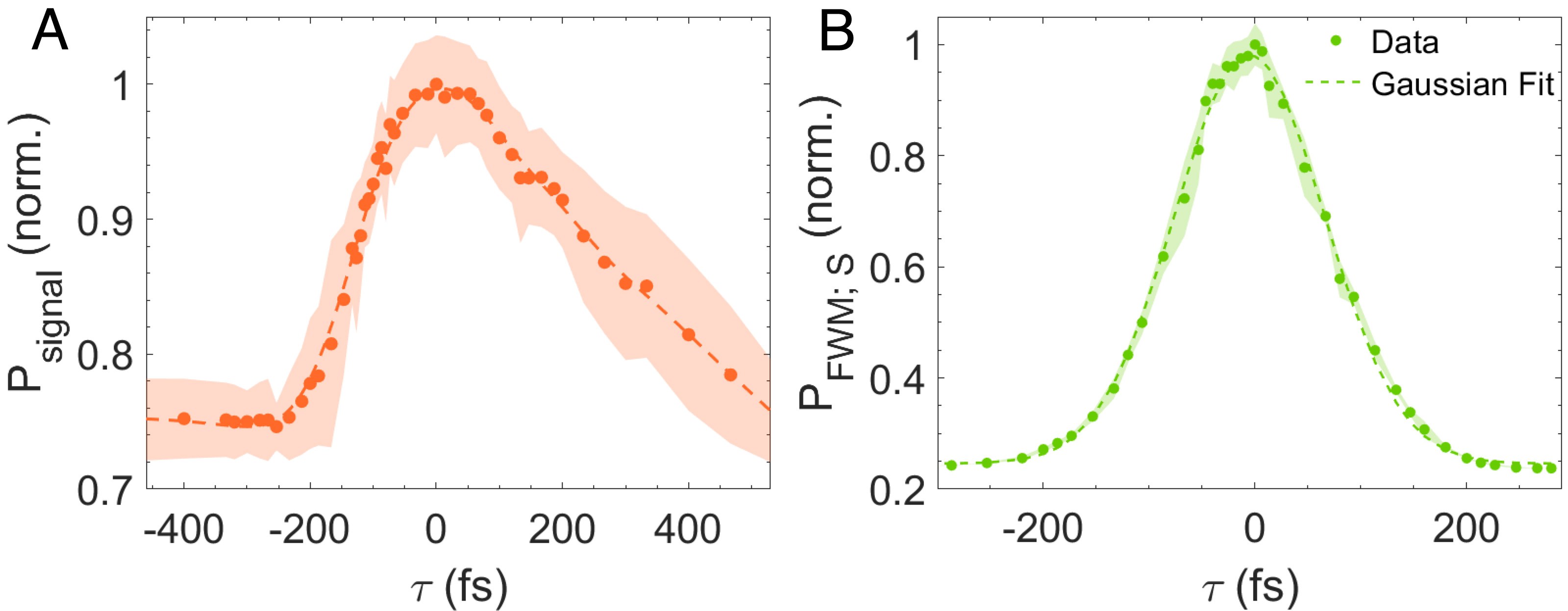}
    \caption{\textbf{The time scales of two nonlinear processes in ITO close to the ENZ wavelength. } The variation of the power of the \textbf{(A)} transmitted signal due to intensity-dependent change in refractive index of ITO, and the \textbf{(B)} sum-FWM response with the temporal delay between the gate and signal pulses. The vertical axes in each plot are normalized to the respective maximum powers at zero temporal delay. }
    \label{fig:ito}
    \end{figure}

We compare the time scales of the two types of nonlinearity of ITO at ENZ wavelength by measuring their nonlinear responses as a function of the time delay between the gate and the probe pulse. The sample used in the experiment is a 310-nm-thick ITO deposited on sapphire, and as shown in the spectra of its permittivity in Fig.~\ref{fig:ito_epsilon}, has an ENZ wavelength of approximately 1515 nm. Figure \ref{fig:ito}A shows the temporal characteristics of the optically induced nonlinear refractive index change caused by thermally excited electrons. The curve of the measured transmitted signal energy shows a steep rising edge, a sign of the fast-rising electron temperature, and a flatter falling edge that corresponds to the electron relaxation. The mechanism of this slow nonlinear effect at ENZ wavelength can be phenomenologically described by the two-temperature model (TTM) \cite{alam2016large, conforti2012derivation}. On the other hand, the sum-four-wave mixing (FWM; S) response shown in Fig. \ref{fig:ito}B does not show any asymmetric temporal features and it agrees with the temporal correlation function between the gate and the signal pulses. The curve is measured by tracing the average power of the generated sum-FWM signal at 517 nm as a function of the gate-signal delay. The gate and signal pulse have energies of 4.3 $\mu$J and 12 $\mu$J respectively, and they are focused onto the ITO with spot sizes of 0.66 mm and 0.24 mm. We note that the signal spot size was approximately 3 times larger in these gate-signal traces than in the imaging measurements due to different lenses used in the two cases for the signal. The average power of the sum-FWM signal reaches a maximum of 50 nW at zero path delay, corresponding to an overall conversion efficiency $\eta = P_{\rm FWM; \, S}/P_{\text{signal}} = 1.25 \times 10^{-5}$, which agrees with the measured efficiency of third-order nonlinear process on ENZ materials \cite{jaffray2022near}. The width of the response curve is primarily limited by the pulse width of the incident gate and signal since the fast electronic nonlinearity is instantaneous. Therefore, the FWM time gating should lead to a better rejection of the scattering effect in our experiment.

\subsubsection*{OAM sorting with cylindrical lenses}
\begin{figure}
    \centering
    \includegraphics[width=\textwidth]{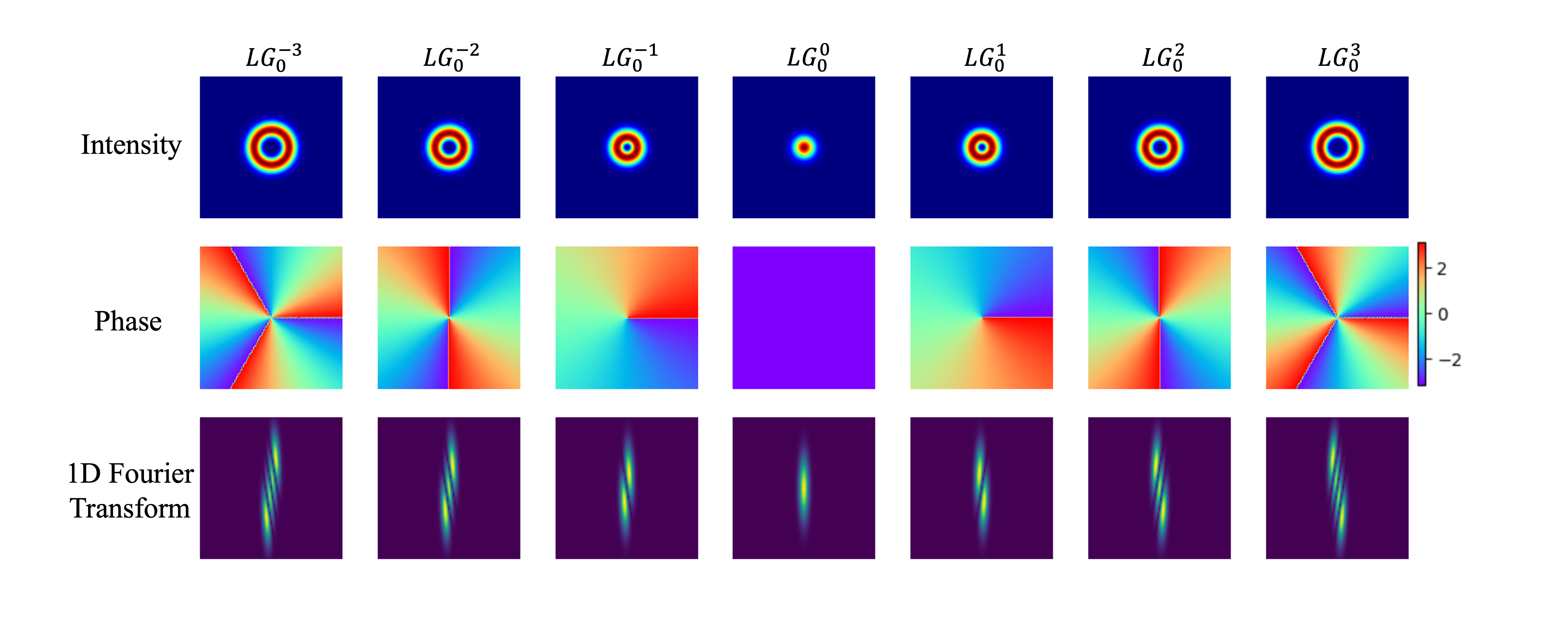}
    \caption{\textbf{Amplitude and phase structures of various Laguerre-Gaussian modes.} Intensity (top row) and phase (middle row) profiles of example Laguerre-Gaussian (LG) modes with $-3 \leq l \leq 3$. The bottom row shows the intensity profiles of the respective LG modes at the focal plane of a single cylindrical lens. The orientation and the number of minima are related to the OAM carried by the incident beam. }
    \label{fig:oam_spatial}
    \end{figure}

To verify that the OAM carried by the signal field is transferred to the sum-FWM field after propagating through scattering media, we use a stationary cylindrical lens and a camera to measure the OAM carried by the two fields \cite{alperin2016quantitative}. This method implements a 1D Fourier transform on the incident light through the use of a single cylindrical lens. The OAM of the incident field is determined by the intensity pattern at the focal plane of the cylindrical lens. In this representative example, we use Laguerre-Gaussian (LG) modes as our OAM-carrying beams instead of the actual field distributions obtained after the spiral-phase plate for simplicity. Figure \ref{fig:oam_spatial} shows the intensity profiles (top row) and the phase profiles (middle row) of various LG modes, $\text{LG}_p^l$, with different OAM ($(l)$) values and $p = 0$. The bottom row shows the intensity profiles of the transformed vortex beams at the focal plane of the cylindrical lens. The cylindrical lens performs a 1D Fourier transform along the horizontal axis. The number and the orientation of the minima in the intensity profiles after the cylindrical lens transform indicate the magnitude of the angular mode index, $\left|l\right|$, and the sign of the OAM charge. 

\begin{figure}
    \centering
    \includegraphics[width=0.9\textwidth]{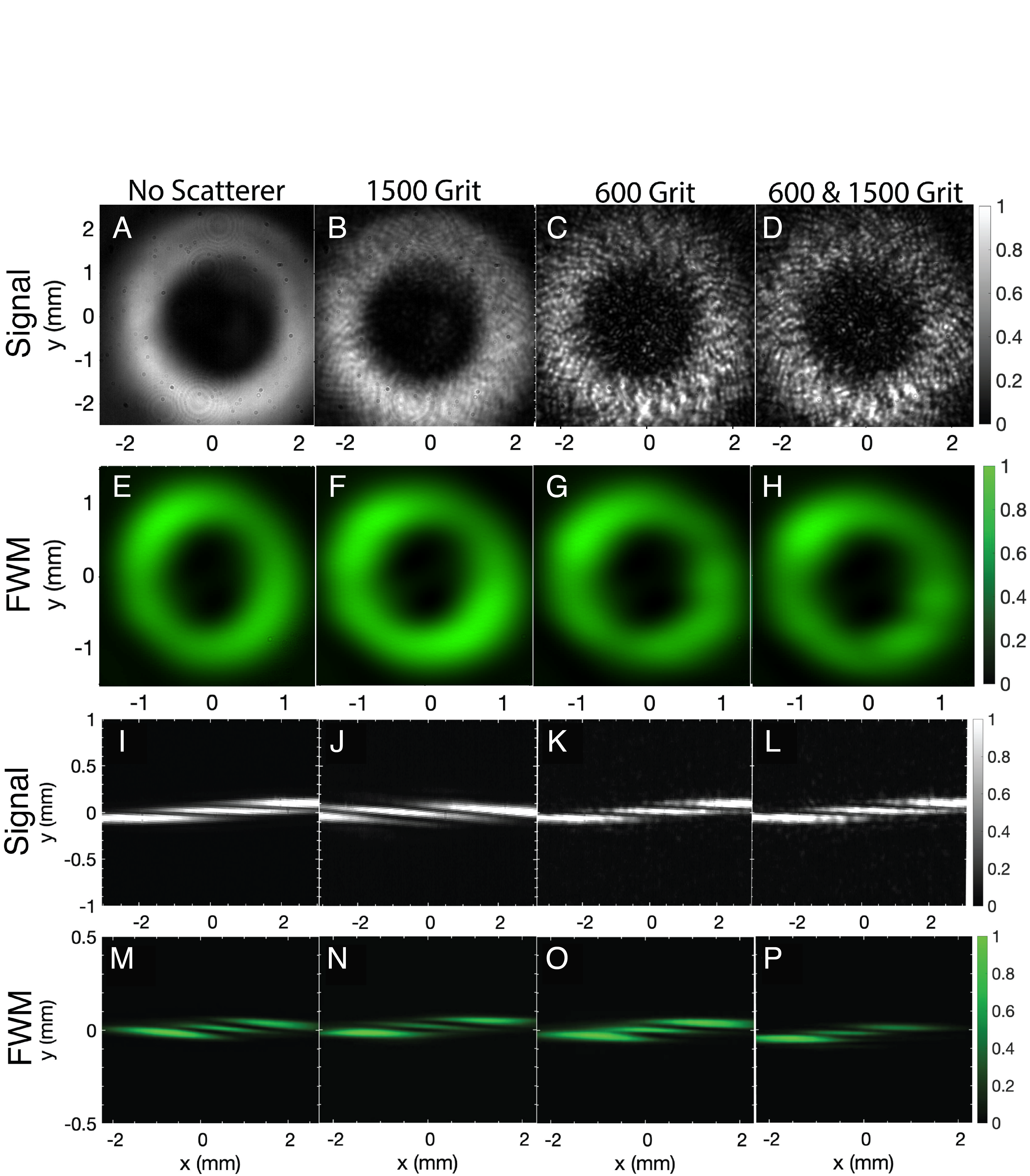}
    \caption{\textbf{Amplitude and phase structures of the OAM-carrying signal and of the generated sum-FWM pulses after optical diffusers with different scattering strength.} Spatial profiles of the \textbf{(A)-(D)} incident signal field carrying an OAM of l = 2 and of the \textbf{(E)-(H)} sum-FWM field with different levels of scattering specified at the top. \textbf{(I)-(L)} The corresponding spatial profiles at the focus of a cylindrical lens of the scattered OAM-carrying signal and \textbf{(M)-(P)} the spatial profiles of the sum-FWM at the focal plane of a cylindrical lens. }
    \label{fig:oam_scattering}
    \end{figure}

Figures \ref{fig:oam_scattering}A-D show the spatial profiles of the OAM-carrying signal field collected by the IR camera (Bobcat 640) under different scattering conditions. Figures \ref{fig:oam_scattering}I-L show the corresponding intensity patterns at the focal plane of the cylindrical lens, and indicate that the signal field predominantly carries a topological charge $l$ of 2 in the presence of scattering. However, both the spatial profiles and the Fourier-transformed patterns show an increasing amount of noise with increased scattering, thereby indicating a decrease in the purity of the OAM state. On the contrary, the spatial profiles (Figs. \ref{fig:oam_scattering}E-H) and the Fourier-transformed patterns (Figs. \ref{fig:oam_scattering}M-P) at the focal plane of the cylindrical lens in the sum-FWM arm stay free from scattering noise at all levels of scattering. 

\subsubsection*{Removal of the effect of time-varying scattering}
\begin{figure}
    \centering
    \includegraphics[width=0.9\textwidth]{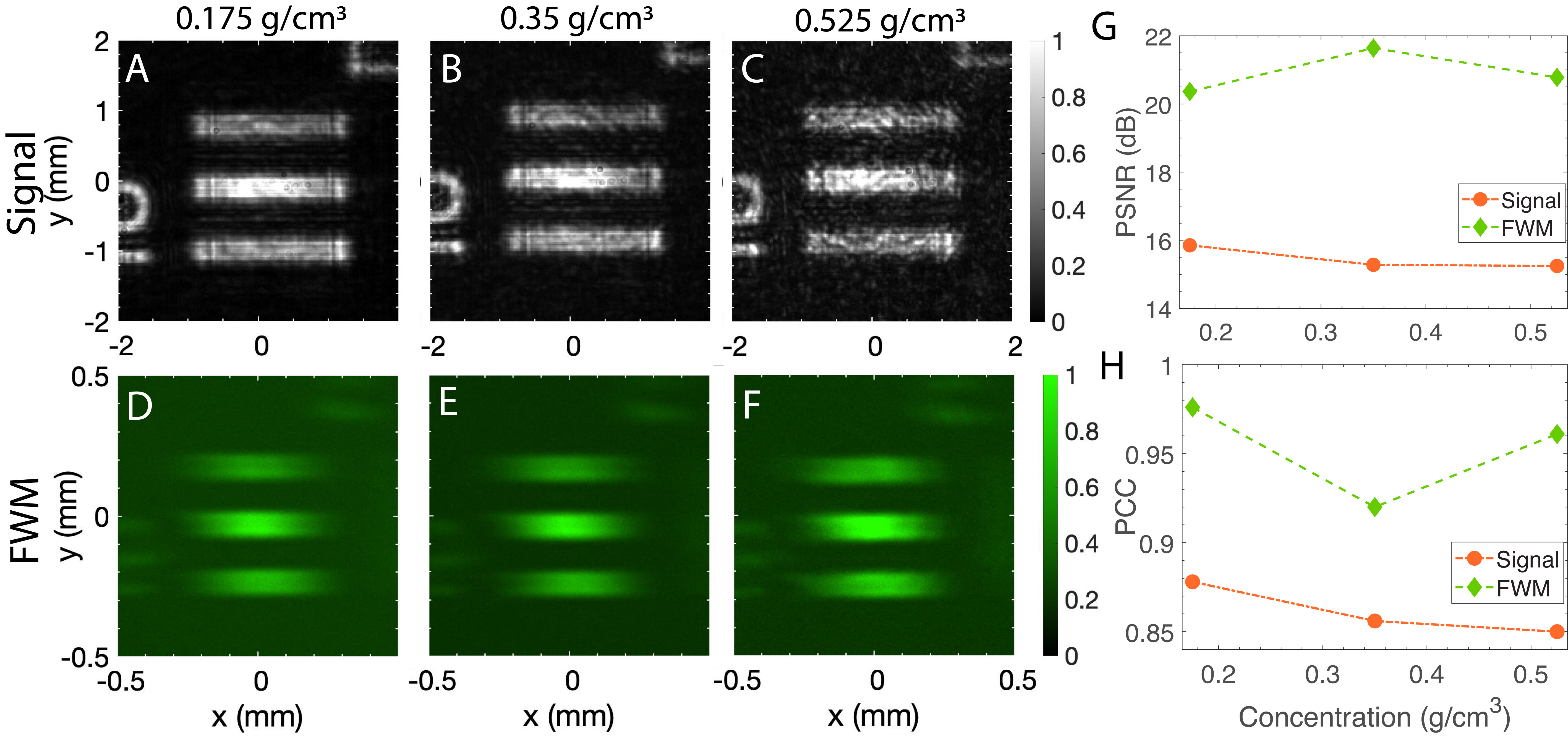}
    \caption{\textbf{Freeze frames of the direct signal and the gated sum-FWM images with different concentrations of polystyrene beads.} \textbf{(A)-(C)} Direct signal images and \textbf{(D)-(F)} sum-FWM images of the USAF target, and the \textbf{(G)} peak signal-to-noise ratio (PSNR) and \textbf{(H)} Pearson correlation coefficient (PCC) of the images for various bead concentrations. }
    \label{fig:latex_beads}
    \end{figure}

We now compare the freeze frames of the scattered signal image and the gated sum-FWM image of the USAF target for polystyrene concentrations ranging from 0.175 g/cm$^3$ to 0.525 g/cm$^3$. The image quality of the direct signal (Fig. \ref{fig:latex_beads}A-C) degrades as the concentration of latex beads increases, which is also reflected in the PSNR and PCC metrics shown in Figs. \ref{fig:latex_beads}G and H, respectively. However, the sum-FWM images (Fig. \ref{fig:latex_beads}D-F) remain free of any scintillation noise. 

\begin{figure}
    \centering
    \includegraphics[width=\textwidth]{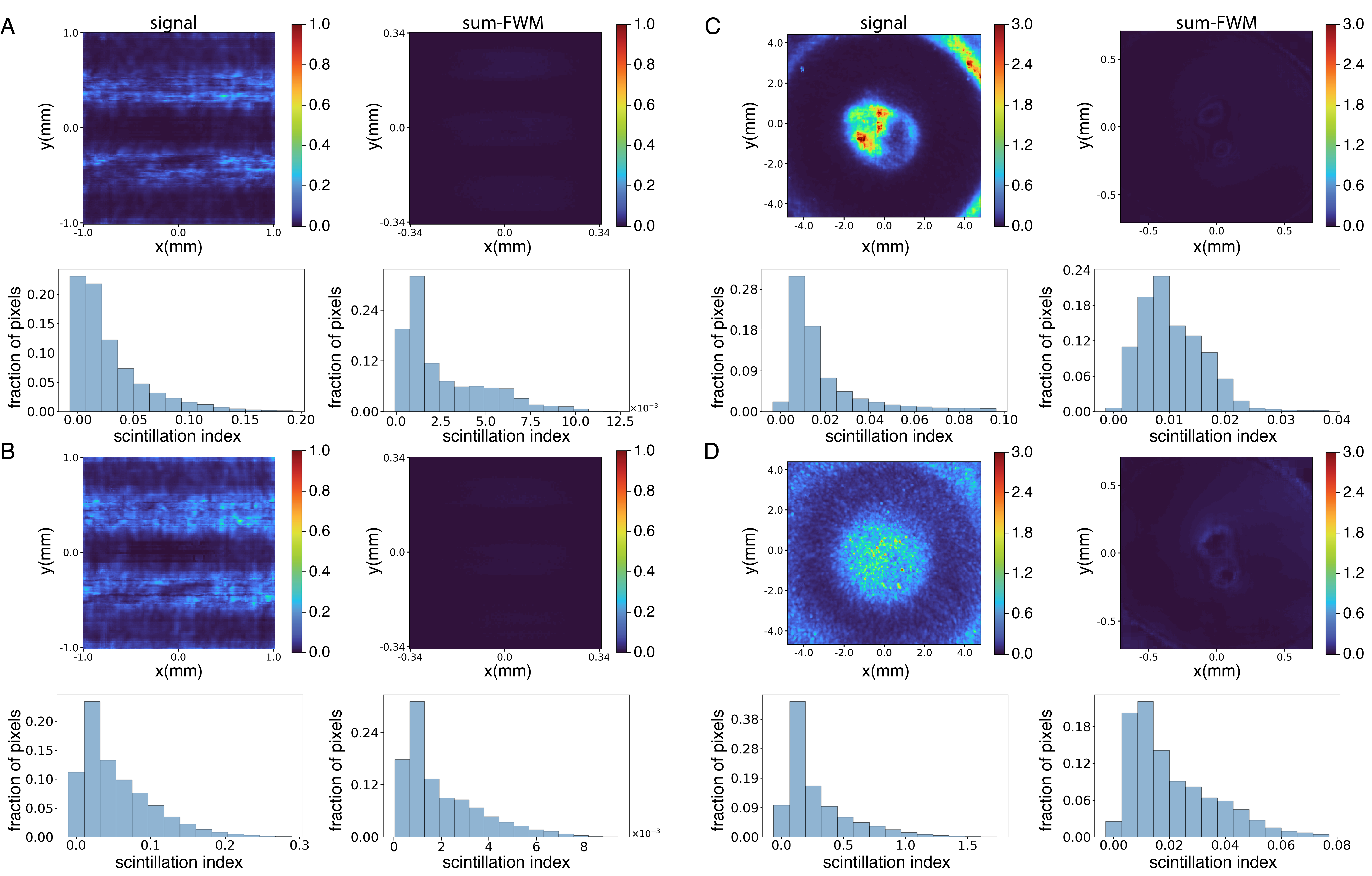}
    \caption{\textbf{Suppression of scintillation due to dynamic scattering by the time-gate.} \textbf{(A), (B)} Scintillation maps and the associated histogram counts of the USAF target in the direct signal images (left column) and the sum-FWM images (right column). The scattering medium is an aqueous suspension of polystyrene beads with a mass concentration of \textbf{(A)} 0.175 g/cm$^3$ and \textbf{(B)} 0.350 g/cm$^3$, respectively. \textbf{(C), (D)} Scintillation maps and the associated histogram counts of the OAM beam ($l=2$) in the direct signal images (left column) and the sum-FWM images (right column). The signal is scattered from a 2-mm thick ground glass diffuser. The grit numbers of the diffusers are \textbf{(C)} 1500  and \textbf{(D)} 600, respectively. }
    \label{fig:scint_latex_beads}
    \end{figure}

Figure \ref{fig:scint_latex_beads} shows the scintillation metrics of the direct and FWM images from the bead suspension (left panels) and the rotating ground glass (right panels), which act as time-varying scattering media. The scintillation index at each pixel, defined in Equation \ref{eq:scint_index} of the main text, reflects the fluctuations in intensity at the pixel in time. Figures \ref{fig:scint_latex_beads}A and B show the scintillation maps and the corresponding histogram counts of the USAF target in for bead concentrations of 0.175 g/cm$^3$ and 0.350 g/cm$^3$, respectively. The scattered signal images (left column) show a stronger scintillation effect than the gated sum-FWM images. Figures \ref{fig:scint_latex_beads}C and D show the scintillation maps and the associated histogram counts of the OAM-encoded signal pulse scattered by a single ground glass diffuser with grit numbers of 1500 and 600, respectively. A similar overall reduction in the scintillation indices is observed in the sum-FWM images for both diffusers.

\clearpage 

\paragraph{Caption for Movie S1.}
\textbf{Real-time imaging of the scattered signal and the gated sum-FWM in aqueous suspension of polystyrene beads with a mass concentration of 0.175 g/cm$^{3}$.}
Animated version of Fig.~\ref{fig:latex_beads}A and Fig.~\ref{fig:latex_beads}D.

\paragraph{Caption for Movie S2.}
\textbf{Real-time imaging of the scattered signal and the sum-FWM in aqueous suspension of polystyrene beads with a mass concentration of 0.525 g/cm$^{3}$.}
Animated version of Fig.~\ref{fig:latex_beads}C and Fig.~\ref{fig:latex_beads}F.

\paragraph{Caption for Movie S3.}
\textbf{Real-time imaging of the scattered OAM-carrying signal pulse and the sum-FWM through ground glass optical diffusers.}
Animated version of Fig.~\ref{fig:oam_images}B and Fig.~\ref{fig:oam_images}F.



\end{document}